\documentclass[final]{jfm}

\makeatletter
\gdef\@underjournal{}
\makeatother

\makeatletter
\gdef\@journal{}
\makeatother

\makeatletter
\def\ps@titlepage{}
\makeatother

\usepackage{graphicx}
\usepackage{newtxtext}
\usepackage{newtxmath}
\usepackage{natbib}
\usepackage{hyperref}
\hypersetup{
	colorlinks = true,
	urlcolor   = blue,
	citecolor  = black,
}

\newcommand{\RomanNumeralCaps}[1]
\usepackage{graphicx,amssymb,amsmath,epsfig,soul,rotating,overpic,varwidth,xcolor}

\usepackage[export]{adjustbox}
\usepackage{subfigure}
\usepackage{subfigmat}
\usepackage{algorithm}
\usepackage[noend]{algpseudocode} 
\usepackage{bm}
\usepackage{xfrac}
\usepackage{multicol}
\usepackage{multirow}
\usepackage{pbox}
\usepackage{tabularx}
\usepackage{setspace}
\usepackage{booktabs}
\usepackage{tikz}
\usetikzlibrary{tikzmark}

\usepackage{lipsum}

\usepackage{color}
\definecolor{blue}{rgb}{0, 0.4470, 0.7410}
\definecolor{red}{rgb}{0.8500, 0.1250, 0.0480} 
\definecolor{green}{rgb}{0.4660, 0.6740, 0.1880}
\definecolor{blue2}{rgb}{0, 0.4470, 0.7410}
\definecolor{Orange}{cmyk}{0 0.6 1 0} % added
\definecolor{red2}{rgb}{0.8500, 0.1250, 0.0480} % added

\newcommand{\bb}[1]{\textbf{\textsf{#1}}}

\begin{document}
	
	% \captionsetup{font=scriptsize,labelfont=scriptsize}

	\shorttitle{Optimally time-dependent modes of vortex gust-airfoil interactions} 
	\shortauthor{Y. Zhong et al.} 
	
	\title{Optimally time-dependent modes of vortex gust-airfoil interactions} 
	
	\author
	{
		Yonghong Zhong\aff{1}\corresp{yhzhong@g.ucla.edu},
        Alireza Amiri-Margavi\aff{2},
        Hessam Babaee\aff{2},
        \and Kunihiko Taira\aff{1}
	}

	\affiliation
	{
		\aff{1}Department of Mechanical and Aerospace Engineering, University of California, Los Angeles, CA 90095, USA
        \aff{2}Department of Mechanical Engineering and Materials Science, University of Pittsburgh, Pittsburgh, PA 15260, USA
		
	}

	\maketitle
	
\begin{abstract}

We find the optimally time-dependent (OTD) orthogonal modes about a time-varying flow generated by a strong gust vortex impacting a NACA 0012 airfoil. 
This OTD analysis reveals the amplification characteristics of perturbations about the unsteady base flow and their amplified spatiotemporal structures that evolve over time.  
We consider four time-varying laminar base flows in which a vortex {\color{black}with a} strength corresponding to the gust ratio $G$ of $\{-1,-0.5,0.5,1\}$ impinges on the leading edge of the airfoil at an angle of attack of $12^\circ$.
In these cases, the impingement of the strong gust vortex causes massive separation and the generation of large-scale vortices around the airfoil within two convective time units.  
As these flow structures develop around the airfoil on a short time scale, the airfoil experiences large transient vortical lift variations in the positive and negative directions that are approximately five to ten times larger than the baseline lift.
The highly unsteady nature of these vortex-airfoil interactions necessitates an advanced analytical technique capable of capturing the transient perturbation dynamics.
For each of the considered gust ratios, the OTD analysis identifies the most amplified region to perturbations, the location of which changes as the wake evolves differently. 
For interactions between a moderate positive vortex gust ($G=0.5$) and the airfoil, the area where perturbations are amplified transitions from the leading-edge vortex sheet to the forming leading-edge vortex. Later, this most amplified structure becomes supported in the airfoil wake directly behind the trailing edge. In contrast, a strong vortex gust ($G=\pm 1$) encountered by the airfoil shows the most amplified OTD mode to appear around the core of the shed vortices.
This study provides an analysis technique and fundamental insights into the broader family of unsteady aerodynamic problems.

\end{abstract}

\section{Introduction}
\label{sec:intro}

Flying in a gusty environment is challenging because wings can experience massive flow separation and violent transient force fluctuations.
One critical gust parameter is the relative velocity between the gust and the freestream, often referred to as the gust ratio $G$~\citep{jones2020gust}. The gust ratio indicates how strong a gust is{\color{black};} a larger-amplitude gust is likely to incite more intense flow {\color{black}efforts on a flying body}.
In nature, the profile and strength of aerodynamic disturbances can vary significantly depending upon the environment of interest{\color{black}. For these reasons, previous studies have considered transient gusts, including streamwise gusts, transverse gusts, and vortex gusts. Among these various types of gusts, a spanwise vortex-gust elicits perhaps the most fluctuations in lift.}
When such a gust interacts with an airfoil, the resulting dynamics are transient and nonlinear, posing considerable {\color{black}challenges} in understanding the unsteady flow behavior. 
However, it is crucial to understand such effects that large-amplitude vortex gusts pose on flying bodies.

Unsteady aerodynamic models have been developed to capture the {\color{black}dominant} effects of gust-airfoil interaction.
For small-amplitude gust encounters where attached flow is assumed around the airfoil, linear thin airfoil theory, and its extensions have been used to build theoretical models for both discrete \citep{von1938airfoil,sears1941some} and periodic \citep{atassi1984sears} gust disturbances. 
In fact, linear aerodynamic models such as the K$\ddot{\rm u}$ssner's model have been known to be effective for a variety of disturbances, including cases where gust-induced flow separation occurs \citep{kussner1936zusammenfassender,badrya2021effect}.
However, linear models become less accurate when the gust ratio is larger than 0.5 or when the angle of attack is higher than $20^\circ$ in which case nonlinear effects are significant.  
While advancements in theoretical and analytical models have been made, characterizing the detailed transient gust-airfoil interactions in a global manner remains elusive due to the high dimensionality of the full-order problem.

To capture the dynamics of global flow fields, modal analysis is a useful tool that extracts dominant features of high-dimensional flows.
For example, proper orthogonal decomposition (POD) \citep{lumley1967structure,aubry1988dynamics} identifies spatially energetic modes, and dynamic mode decomposition (DMD) \citep{schmid_2010} {\color{black}extracts} spatial structures that are associated with the spectral content of flow dynamics.
Modal analysis can also reveal the stability and transition characteristics of fluid flows.  
Global stability analysis~\citep{theofilis2011global} reveals the dominant stability modes from the linearized Navier--Stokes equations about a given steady state. The linear eigenvalues found from the global stability analysis provide information about the growth or decay rates of perturbations with respect to the base flow.
Based on an energy measure for the time-dependent response of the flow~\citep{schmid2007nonmodal}, the non-modal analysis framework can evaluate energy amplification by analyzing the harmonic response from harmonic forcing inputs~\citep{farrell1993stochastic} and formulate an initial-value problem to examine the transient energy growth over a finite time interval~\citep{blackburn2008convective}.
Resolvent analysis \citep{trefethen1993hydrodynamic, jovanovic2005componentwise} is a method for understanding the response of a dynamical system to external forcing or perturbations. The response and forcing modes obtained from the resolvent analysis reveal the most amplified perturbation structures and how they are excited by external forcing.
The resolvent analysis was extended for turbulent flows by treating the nonlinearity in the perturbation equation as forcing~\citep{mckeon2010critical}. The coherent structures identified by resolvent analysis provide physical interpretations in wall turbulence~\citep{moarref2013model} and turbulent cavity flows~\citep{gomez2016reduced}. These modal techniques can provide fundamental behavior of the flow response, which can help understand turbulent flows and implement flow control \citep{yeh_taira_2019,LiuJFM2021}.

Although the aforementioned modal analysis techniques offer valuable insights into the characterization of complex fluid flows \citep{taira2017modal, taira2020modal, UnniPAS2023}, {\color{black}the majority} of these methods are built on the assumption of a time-invariant base flow. These analysis techniques require careful generalization for analyzing unsteady base flows. To address this issue, optimally time-dependent (OTD) mode decomposition \citep{babaee2016minimization} has been developed, where linear stability analysis can be performed for perturbation growth around arbitrarily time-dependent base flows. In particular, the OTD analysis is applied to the instantaneously linearized dynamics, and an evolution equation is derived for a set of orthonormal time-dependent modes. It has been shown that the OTD modes converge exponentially fast to the dominant eigenvectors associated with the largest finite-time Lyapunov exponents \citep{babaee2017reduced}. The OTD mode decomposition is closely related to dynamical low-rank approximation \citep{KL07} and dynamically orthogonal decomposition \citep{SL09}. The OTD method uncovers the time-dependent orthonormal modes that capture the dominant transient amplification of perturbations with respect to time-varying base flows, hence serves as a powerful tool to characterize the perturbation dynamics of various unsteady flows \citep{kern_hanifi_henningson_2022,donello2022computing,Beneitez_M_2023,amirimargavi2023lowrank,Kern_Negi_Hanifi_Henningson_2024}. 

In this study, we employ OTD mode analysis to investigate vortex-airfoil interaction{\color{black}s}, where the vortex gust ratio {\color{black}magnitude} exceeds 0.5. The complexity of such interactions, characterized by strong nonlinearities, transient dynamics, and high-dimensional flow structures, demands a method that can adaptively capture perturbation dynamics of the evolving flow field. OTD modes, which are found at each time step, offer a dynamic approach {\color{black}for} dimensionality {\color{black}reduction} and reveal the optimal timing and locations of critical perturbation amplifications. This makes OTD mode analysis particularly suited for studying the violent aerodynamic phenomena associated with large vortex gusts, where traditional methods fall short in capturing the essential dynamics.

By capturing the evolving flow structures and identifying regions of sensitivity at precise moments in time, OTD mode analysis enables the possibility of designing highly effective, time-varying flow control strategies. This dynamic approach to flow manipulation is critical for optimizing aerodynamic performance and mitigating adverse flow effects. 
For example, Blanchard and Sapsis~\citep{blanchard2019stabilization} proposed a strategy for identifying the optimal control domain using a criterion derived from the OTD modes. They demonstrated that OTD-based control can successfully alleviate the flow unsteadiness by guiding flow trajectories toward the desired fixed point.

{\color{black}The current} study is organized as follows. We start section~\ref{sec:methodology} {\color{black}by} presenting the methodology of OTD analysis. The model problem of vortex-airfoil interactions and the related physics are described in section~\ref{sec:setup}. In section~\ref{sec:results}, the amplifications of perturbations for four vortex-airfoil interaction cases are examined with the OTD analysis. Finally, the conclusions are presented in section~\ref{sec:conclusion}.

\section{Methodology: optimally time-dependent mode analysis}
\label{sec:methodology}

Our objective is to find the dominant transient amplification mechanisms of non-periodic time-dependent flows using the optimally time-dependent (OTD) mode analysis~\citep{babaee2016minimization}.
To identify the transient amplification of perturbations about an unsteady vortex gust-airfoil interactions, we consider the flow state ${\textsf{\textbf q}}(t)$ to be comprised of the base flow (trajectory) ${\bar{\textsf{\textbf q}}}(t)$ and perturbation ${\textsf{\textbf q}}^{\prime}(t)$,
\begin{equation}
    {\textsf{\textbf q}}(t)
    ={\bar{\textsf{\textbf q}}}(t)+{\textsf{\textbf q}}^{\prime}(t) \in {\mathbb{R}}^{n}.
\end{equation}
Here, it is assumed that the flow domain is spatially discretized{\color{black}, and} $n$ is the degrees of the freedom of the discretized flow, i.e., the number of grid points times the number of state variables $[\rho,\rho u, \rho v, \rho w, e]$.
We substitute this flow expression into the Navier-Stokes equations to derive the linear evolution equation for the perturbation ${\textsf{\textbf q}}^{\prime}(t)$ about an arbitrary trajectory ${\bar{\textsf{\textbf q}}}(t)$.  With the assumption that the perturbation magnitude is small, we find that 
\begin{equation}
    \frac{d \textbf{\textsf{q}}^{\prime}(t)}{dt}
    =\textbf{\textsf{L}}_{{\bar{\textsf{\textbf q}}}}(t)\textbf{\textsf{q}}^{\prime}(t).
    \label{eq:pert}
\end{equation}
The time-varying linear operator $\textbf{\textsf{L}}_{{\bar{\textsf{\textbf q}}}}(t)\in {\mathbb{R}}^{n\times n}$ is derived about the unsteady base state ${\bar{\textsf{\textbf q}}}(t)$. 
For convenience, we drop the subscript from  $\textbf{\textsf{L}}_{{\bar{\textsf{\textbf q}}}}(t)$ and denote it as $\textbf{\textsf{L}}(t)$.

In this paper, we take the base state ${\bar{\textsf{\textbf q}}}(t)$ to be the unsteady flow produced by a gust vortex impinging on an airfoil. Through the present analysis, we aim to identify the dominant temporary evolving spatial structures susceptible to amplification during the vortex-airfoil interaction. The dynamics of the unsteady base flows will be discussed in section~\ref{sec:physics}.

Now, we consider a collection of $d$ initial perturbations 
\[
\textbf{\textsf{Q}}^{\prime}(t_0)\equiv[\textbf{\textsf q}^{\prime}_{1}(t_0),\textbf{\textsf q}^{\prime}_{2}(t_0),...,\textbf{\textsf q}^{\prime}_{d}(t_0)]
\in {\mathbb{R}}^{n\times {d}}
\]
and evolve them over time using linear equation~\ref{eq:pert} such that 
\begin{equation}
    \frac{d \textbf{\textsf{Q}}^{\prime}(t)}{dt}
    =\textbf{\textsf{L}}(t)\textbf{\textsf{Q}}^{\prime}(t).
    \label{eq:pert_set}
\end{equation}
The aim here is to take the collection of perturbation trajectories $\textbf{\textsf{Q}}^{\prime}(t)$ that dynamically evolve about the time-varying base state ${\bar{\textsf{\textbf q}}}(t)$ and determine the dominant modes that capture their time-varying amplification characteristics.
To approximate the perturbations $\textbf{\textsf{Q}}^{\prime}(t)$ in a reduced-order subspace, {\color{black}we take a low-rank approximation~\citep{babaee2016minimization,RAMEZANIAN2021113882} as} 
\begin{equation}
        \textbf{\textsf{Q}}^{\prime}(t)\approx \textbf{\textsf{U}}_r(t)\textbf{\textsf{Y}}_r(t)^{\rm T},
        \label{eq:linear}
\end{equation}  
where 
\begin{equation}
    \textbf{\textsf{U}}_r(t)
    \equiv[\textbf{\textsf u}_1(t), \textbf{\textsf u}_2(t),
    ...,\textbf{\textsf u}_r(t)]
    \in {\mathbb{R}}^{n\times r}
    \label{eq:U_stacked}
\end{equation}        
is a set of $r$ time-dependent orthonormal basis vectors, and 
\begin{equation}
    \textbf{\textsf{Y}}_r(t)\equiv[\textbf{\textsf y}_1(t), \textbf{\textsf y}_2(t),...,\textbf{\textsf y}_r(t)]\in {\mathbb{R}}^{d\times r}
\end{equation}
is the reduced-order coefficient matrix. {\color{black} The full-order perturbation dynamics can be approximated properly with an appropriate choice of $r$ basis vectors and their coefficients.}

% Here, $r$ is taken to be $r \ll d$. {\color{black}With a limited $r$ number of $\textbf{\textsf{U}}_r(t)$ and $\textbf{\textsf{Y}}_r(t)$,  the low-rank approximation can be a good representative of the full space~\citep{RAMEZANIAN2021113882}.}

To derive the evolution equations for $\textbf{\textsf{U}}_r(t)$ and $\textbf{\textsf{Y}}_r(t)$, we trace the perturbation dynamics by substituting equation~\ref{eq:linear} into the linearized equation~\ref{eq:pert_set}:
\begin{equation}\frac{d(\textbf{\textsf{U}}_r\textbf{\textsf{Y}}_r^{\rm T})}{dt}=\textbf{\textsf{U}}_r\frac{d\textbf{\textsf{Y}}_r^{\rm T}}{dt}+\frac{d\textbf{\textsf{U}}_r}{dt}\textbf{\textsf{Y}}_r^{\rm T} {\color{black}=} \textbf{\textsf{L}}\textbf{\textsf{U}}_r\textbf{\textsf{Y}}_r^{\rm T}{\color{black}.}
    \label{Eq:linearOTD}
\end{equation}
{\color{black}Strictly speaking,} $d(\textbf{\textsf{U}}_r\textbf{\textsf{Y}}_r^{\rm T})/dt \neq \textbf{\textsf{L}}\textbf{\textsf{U}}_r\textbf{\textsf{Y}}_r^{\rm T}$ due to the OTD low-rank approximation error.
The evolution equations for $\textbf{\textsf{U}}_r(t)$ and $\textbf{\textsf{Y}}_r(t)$ can be obtained via a variational principle according to \citet{donello2022computing}, which involves minimizing the residual
{\color{black}\begin{equation}\label{eq:var_prin}
F(\frac{d\textbf{\textsf{Y}}_r}{dt},\frac{d\textbf{\textsf{U}}_r}{dt})=\bigg \| \textbf{\textsf{U}}_r\frac{d\textbf{\textsf{Y}}_r^{\rm T}}{dt}+\frac{d\textbf{\textsf{U}}_r}{dt}\textbf{\textsf{Y}}_r^{\rm T} -  \textbf{\textsf{L}}\textbf{\textsf{U}}_r\textbf{\textsf{Y}}_r^{\rm T} \bigg \|_F,
\end{equation}}
subject to the orthonormality constraints of OTD modes. In the above equation, $\| \cdot \|_F$ is the matrix {\color{black}Frobenius norm}. The first-order optimality conditions of the above minimization principle yield the OTD evolution equations. The individual evolution equations for $\textbf{\textsf{U}}_r$ and $\textbf{\textsf{Y}}_r$ can alternatively be found via the Galerkin projection of equation \ref{Eq:linearOTD} onto $\textbf{\textsf{U}}_r$ and $\textbf{\textsf{Y}}_r$ and enforcing the orthonormality condition of the time-dependent modes $\textbf{\textsf{U}}_r^{\rm T}\textbf{\textsf{U}}_r=\textbf{\textsf{I}}$. Taking the time derivative of the orthonormality condition leads to
\begin{equation}
        \textbf{\textsf{U}}_r^{\rm T}\frac{d\textbf{\textsf{U}}_r}{d t}+\frac{d\textbf{\textsf{U}}_r^{\rm T}}{d t}\textbf{\textsf{U}}_r=\textbf{\textsf{0}},
    \label{eq:ortho}
\end{equation}
where we henceforth denote $\bm{\Phi}=\textbf{\textsf{U}}_r^{\rm T}d \textbf{\textsf{U}}_r/{dt} \in \mathbb{R}^{r \times r}$.
In order to satisfy the above equation, $\bm{\Phi}$ needs to be a skew-symmetric matrix, i.e., $\bm{\Phi}^{\rm T}=- \bm{\Phi}$. The evolution equation for $\textbf{\textsf{Y}}_r(t)$ can be found via projecting equation \ref{Eq:linearOTD} onto $\textbf{\textsf{U}}_r$. This is accomplished by multiplying both sides of that equation with $\textbf{\textsf{U}}_r^{\rm T}$ from the left side, which yields:
\begin{equation}
        \frac{d\textbf{\textsf{Y}}_r^{\rm T}}{dt}=(\textbf{\textsf{U}}_r^{\rm T}\textbf{\textsf{L}}\textbf{\textsf{U}}_r-\bm{\Phi})\textbf{\textsf{Y}}_r^{\rm T},
    \label{eq:Yr_Phi}
\end{equation}
where the orthonormality constraints, $\textbf{\textsf{U}}_r^{\rm T}\textbf{\textsf{U}}_r=\textbf{\textsf{I}}$ and $\bm{\Phi}=\textbf{\textsf{U}}_r^{\rm T}d \textbf{\textsf{U}}_r/{dt}$ are utilized. Similarly, the evolution equation for $\textbf{\textsf{U}}_r(t)$ can be found projecting equation \ref{Eq:linearOTD} onto {\color{black}$\textbf{\textsf{Y}}_r$}, which can be performed by multiplying that equation onto $\textbf{\textsf{Y}}_r$ from the right side. This results in
\begin{equation}
        \frac{d\textbf{\textsf{U}}_r}{d t}=\textbf{\textsf{L}}\textbf{\textsf{U}}_r-\textbf{\textsf{U}}_r(\textbf{\textsf{U}}_r^{\rm T}\textbf{\textsf{L}}\textbf{\textsf{U}}_r-\bm{\Phi}),
        \label{eq:Ur_Phi}
\end{equation}    
where equation \ref{eq:Yr_Phi} is utilized for $d\textbf{\textsf{Y}}_r^{\rm T}/dt$ and the two sides of the equations are multiplied by $(\textbf{\textsf{Y}}_r^{\rm T}\textbf{\textsf{Y}}_r)^{-1}$.
With these equations, we can evolve the time-dependent basis and the coefficient matrix given the time-varying linear operator $\textbf{\textsf{L}}(t)$ and the initial conditions.

To further simplify equations~\ref{eq:Ur_Phi} and \ref{eq:Yr_Phi}, we can choose $\bm{\Phi}=\textbf{\textsf{0}}$. The choice of the skew-symmetric matrix $\bm{\Phi}$ is not unique. With different choices of $\bm{\Phi}$, it can be proven that one OTD subspace is equivalent to another OTD subspace after being rotated by an orthogonal rotation matrix~\citep{babaee2016minimization,blanchard2019analytical}. When we choose a simple choice of $\bm{\Phi} = \textbf{\textsf{0}}$, we have 
\begin{equation}
        \frac{d\textbf{\textsf{U}}_r}{d t}=(\textbf{\textsf{I}}-\textbf{\textsf{U}}_r\textbf{\textsf{U}}_r^{\rm T})\textbf{\textsf{L}}\textbf{\textsf{U}}_r,
\end{equation}
where $\textbf{\textsf{I}}-\textbf{\textsf{U}}_r\textbf{\textsf{U}}_r^{\rm T}$ is the orthogonal projector onto the complement of subspace $\textbf{\textsf{U}}_r$~\citep{donello2022computing}. In this case, we arrive at the closed form of OTD evolution equations of
\begin{equation}
        \frac{d\textbf{\textsf{U}}_r}{dt}=\textbf{\textsf{L}}\textbf{\textsf{U}}_r-\textbf{\textsf{U}}_r(\textbf{\textsf{U}}_r^{\rm T}\textbf{\textsf{L}}\textbf{\textsf{U}}_r)
        \label{eq:U_evolve}
\end{equation}
and
\begin{equation}
        \frac{d\textbf{\textsf{Y}}_r^{\rm T}}{dt}=(\textbf{\textsf{U}}_r^{\rm T}\textbf{\textsf{L}}\textbf{\textsf{U}}_r)\textbf{\textsf{Y}}_r^{\rm T}.
        \label{eq:Y_evolve}
\end{equation}    
Based on the above time-dependent equations, we {\color{black}can} evolve $\textbf{\textsf{U}}_r(t)$ and $\textbf{\textsf{Y}}_r(t)^{\rm T}$ {\color{black}given} the initial conditions $\textbf{\textsf{U}}_r(t_0)$ and $\textbf{\textsf{Y}}_r(t_0)^{\rm T}$. 
Here, the choice of initial condition is also not unique. However, the time-dependent modes from two different initial conditions span the same OTD subspace when one initial condition can be transformed into another one~\citep{babaee2017reduced}.

The variational principle introduced in this work differs slightly from the minimization problem presented in \cite{babaee2016minimization}, though the resulting OTD evolution equations are identical. Since the evolution equation for $\textbf{\textsf{U}}_r$ is independent of $\textbf{\textsf{Y}}_r$, as is evident from Eqs. \ref{eq:U_evolve} and \ref{eq:Y_evolve}, it is possible to formulate a variational principle with respect to {\color{black}${d\textbf{\textsf{U}}_r}/{dt}$}, as shown in \cite{babaee2016minimization} and obtain an evolution equation for $\textbf{\textsf{U}}_r$. The evolution equation for $\textbf{\textsf{Y}}_r^{\rm T}$ can be derived by projecting the full-order model {\color{black}of }Eq. \ref{eq:pert} onto $\textbf{\textsf{U}}_r$. However, in the variational principle presented in Eq. \ref{eq:var_prin}, both {\color{black}${d\textbf{\textsf{U}}_r}/{dt}$} and {\color{black}${d\textbf{\textsf{Y}}_r}/{dt}$}  are control variables and minimizing the functional in Eq. \ref{eq:var_prin} yields the evolution equations for $\textbf{\textsf{U}}_r$ and $\textbf{\textsf{Y}}_r$.  The variational principle presented in this work has the advantage of having a simple interpretation{\color{black}. T}he matrix $\textbf{\textsf{U}}_r\textbf{\textsf{Y}}_r^{\rm T}$ is a low-rank approximation of  $\textbf{\textsf{Q}}^{\prime}$ with $\textbf{\textsf{U}}_r$ and $\textbf{\textsf{Y}}_r$ evolving to minimize the residual due to the low-rank approximation error.

%%%%%%%%%%%%%%%%%%%%%%%%%%%%%%%%%%%%
\begin{figure}
    \centering
    \includegraphics[width=0.45\textwidth]{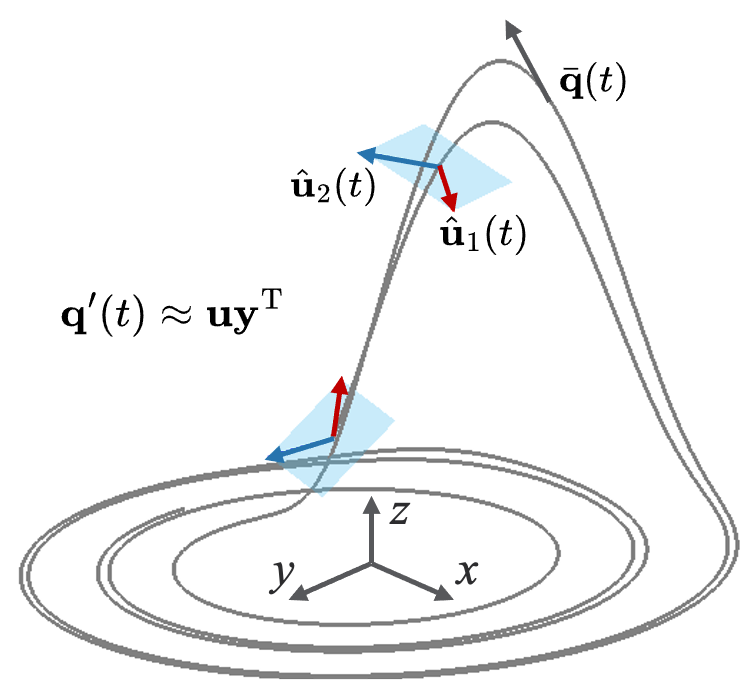}
    \caption{
    The evolution of the base flow ${\bar{\textsf{\textbf q}}}(t)$ and the optimally time-dependent modes $\textbf{\textsf u}_i(t)$ for an example of the R\"ossler system. The perturbation $\textbf{\textsf{q}}^{\prime}(t)$ is captured by the product of optimally time-dependent modes $\textbf{\textsf{u}}(t)$ and their coefficients $\textbf{\textsf{y}}(t)$.}
    \label{OTD}
\end{figure}
%%%%%%%%%%%%%%%%%%%%%%%%%%%%%%%%%%%% 

In the above formulation, the time-dependent modes are correlated with each other. We can perform a rotation so that these modes are independent and ranked by their importance. To this end, let us consider a correlation matrix $\textbf{\textsf{C}}(t)=\textbf{\textsf{Y}}_r^{\rm T}(t) \textbf{\textsf{Y}}_r(t)\in {\mathbb{R}}^{r\times r}$.  The eigenvalue decomposition of this correlation matrix $\textbf{\textsf{C}}(t)$ yields
\begin{equation}
        \textbf{\textsf{C}}(t)\textbf{\textsf{R}}(t)=\textbf{\textsf{R}}(t){\bm\Lambda}(t),
\end{equation}
where $\bm\Lambda (t)\equiv\text{diag}(\lambda_1(t),\lambda_2(t),...,\lambda_\textit{r}(t))$ holds the set of eigenvalues and $\textbf{\textsf{R}}(t)\in {\mathbb{R}}^{r\times r}$ is composed of the corresponding eigenvectors of $\textbf{\textsf{C}}(t)$. Since $\textbf{\textsf{C}}(t)$ is a {\color{black}symmetric} positive matrix, $\textbf{\textsf{R}}(t)$ is an orthonormal matrix, and $\bm\Lambda (t)$ has all nonnegative eigenvalues of $\lambda_1\geq\lambda_2\geq ...\geq \lambda_r$. Here, $\textbf{\textsf{C}}(t)$ is generally a full matrix, which indicates that the coefficients are correlated.

A linear mapping from the correlated coefficients $\textbf{\textsf{Y}}_r(t)$ to the uncorrelated coefficient matrix $\hat{\textbf{\textsf{Y}}}_r(t)\bm\Sigma_r(t)$ can be found {\color{black}using the fact that the perturbations $\textbf{\textsf{U}}_r\textbf{\textsf{Y}}_r^{\rm T}$ can be decomposed through a singular value decomposition} $\hat{\textbf{\textsf{U}}}_r(t)\bm\Sigma_r(t)\hat{\textbf{\textsf{Y}}}_r(t)^{\rm T}$. Such a mapping is realized by performing a matrix rotation and scaling
\begin{equation}        \hat{\textbf{\textsf{Y}}}_r(t)=\textbf{\textsf{Y}}_r(t)\textbf{\textsf{R}}(t){\bm \Sigma_r}^{-1}(t),
\end{equation}
where $\bm\Sigma_r (t)\equiv\rm{diag}(\sigma_1(t),\sigma_2(t),...,\sigma_r(t))$ holds singular values $\sigma_i(t)=\lambda_i(t)^{1/2}$.
Thus, we now have the ranked spatial modes $\hat{\textbf{\textsf{U}}}_r(t)$ based on the eigenvalues $\lambda_i(t)$:
\begin{equation}      \hat{\textbf{\textsf{U}}}_r(t)=\textbf{\textsf{U}}_r(t)\textbf{\textsf{R}}(t).
\end{equation}
We will henceforth refer to this ranked time-dependent modes $\hat{\textbf{\textsf{U}}}_r(t)$ as the {\it{OTD modes}}. 
Since the leading mode possesses the largest singular value, the primary amplified structure of perturbations can be identified from the leading time-dependent mode. 
As an illustrative example, we show the optimally time-dependent modes of three-dimensional R\"ossler system in figure~\ref{OTD}. The {\color{black}two }time-evolving vectors indicate the two {\color{black}dominant amplification} directions of the perturbations.

One of the utilities of the OTD analysis is that it reveals the most amplified initial perturbation and the corresponding instantaneous amplification factor. The first singular vector and singular value of the OTD low-rank approximation contain information about the optimal perturbation. In particular, the most amplified perturbation is represented as $\bb q_{*_1}'(t) = \sigma_1(t) \hat{\bb{u}}_1(t)$ and the optimal initial perturbation that leads to the maximum amplification at time $t$ is obtained via
\begin{equation}\label{eq:inti_opt}
\bb q_{0_1}'(t) =  \hat{\bb{U}}_r(t_0) \bm \Sigma_r(t_0) \hat{\bb Y}_r(t_0)^{\rm T} \hat{\bb y}_1(t).
\end{equation}
The amplitude of the optimal perturbation is $\|\bb q_{*_1}'(t) \|_2 = \|\sigma_1(t) \hat{\bb{u}}_1(t) \|_2=\sigma_1(t)$, since $\| \hat{\bb{u}}_1(t) \|_2=1$.  {\color{black}T}he optimal initial perturbation is also time-dependent {\color{black}as evident from Equation \ref{eq:inti_opt}}. That is, for each choice of $t$, a different optimal perturbation condition achieves the maximum amplification at time $t$ and as time evolves, the optimal initial perturbation $\bb q_{0_1}'(t) $ varies smoothly.
{\color{black}The optimal initial perturbation is confined to the space spanned by the columns of $\hat{\textbf{\textsf{U}}}_r(t_0){\rm\bm \Sigma}_r(t_0)\hat{\textbf{\textsf{Y}}}_r^{\rm T}(t_0)$. }
To obtain the maximum energy amplification, denoted as $g_1(t)$, the amplitude of the perturbation should be normalized with respect to the amplitude of the initial perturbation
\begin{equation}
 g_1(t) \equiv \frac{\| \bb q'_{*_1}(t)\|^2}{\|\bb q_{0_1}'(t) \|^2} =\frac{\sigma^2_1(t)}{\|\bb q_{0_1}'(t) \|^2}{\color{black}.}   
\end{equation}
% where $g_1(t)$ represents the peak amplification that can be achieved among all initial perturbations.

Similarly, the higher-order sub-optimal initial perturbations are 
\begin{equation}\label{eq:inti_subopt}
\bb q_{0_i}'(t) =  \hat{\bb{U}}_r(t_0) \bm \Sigma_r(t_0) \hat{\bb Y}_r(t_0)^{\rm T} \hat{\bb y}_i(t), \quad i=2, 3, ..., r.
\end{equation}
The corresponding amplified perturbations are $\bb q_{*_i}'(t) = \sigma_i(t) \hat{\bb{u}}_i(t)$ for $i=2,3,...,r$.
Therefore, the levels of energy amplification $g_i(t)$ are found to be
\begin{equation}
 g_i(t) \equiv \frac{\| \bb q'_{*_i}(t)\|^2}{\|\bb q_{0_i}'(t) \|^2} =\frac{\sigma^2_i(t)}{\|\bb q_{0_i}'(t) \|^2}, \quad i=2, 3, ..., r{\color{black}.}   
\end{equation}
{\color{black}F}or each OTD mode $\hat{\bb{U}}_i$, $g_1(t)$ is the maximum possible amplification of the perturbation energy that can occur in the fluid system over a given time horizon{\color{black}. T}he higher-order energy amplification $g_i(t)$, $i>1$, may also capture important growth. Analyzing the higher-order energy amplifications provides a {\color{black}supplemental} understanding of the transient dynamics of the fluid system. 
Each singular value $\sigma_i(t)$ indicates the magnitude of each OTD mode, as shown in the SVD form of low-rank approximation of perturbations $\textbf{\textsf{Q}}^{\prime}(t)\approx\hat{\textbf{\textsf{U}}}_r(t)\bm\Sigma_r(t)\hat{\textbf{\textsf{Y}}}_r(t)^{\rm T}$. 
On the other hand, $g_i(t)$ is a direct indicator of how much energy is amplified for each OTD mode with respect to the initial perturbation energy. 

\section{Model Problem}
\label{sec:setup}
\subsection{Problem Setup}

\begin{figure}
  \centering
    \includegraphics[width=1\textwidth]{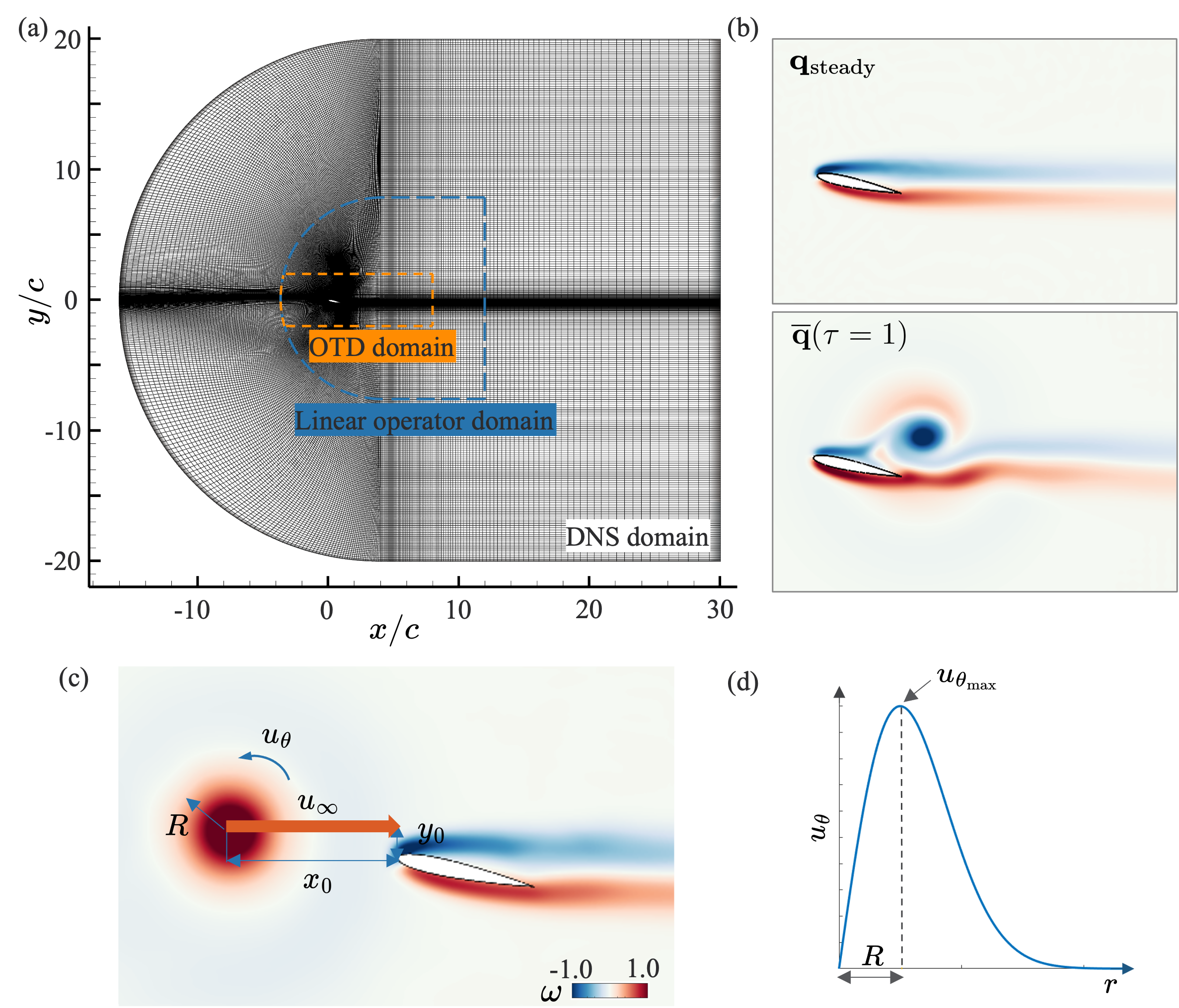}
    \caption{(a) Computational domains of DNS, linear operator, and OTD mode analysis for vortex-airfoil interaction problem.
    (b) Vorticity fields of steady state (without vortex gust) and time-varying base state at ${\color{black}\tau}=1$.
    (c) Parameters of vortex-airfoil interaction problem. (d) Velocity profile of the vortex gust.}
    \label{para}
\end{figure}

The OTD modes identify the transient amplification of a time-varying flow. With strong vortex-airfoil interactions exhibiting violent transient flow features, OTD {\color{black}mode} analysis can be useful for capturing the perturbation dynamics.
We study the transient amplifications of the complex interactions between a vortical gust and the NACA 0012 airfoil, as shown in figure~\ref{para}. The base flow considered in OTD mode analysis is the flow disturbed by a vortex gust. 
Direct numerical simulations (DNS) of flows with and without a gust vortex-impingement are performed with the compressible flow solver CharLES~\citep{khalighi2011noise,khalighi2011unstructured,bres2017unstructured}.
Before interacting with a vortex gust, the two-dimensional flow is steady over an airfoil at an angle of attack of $12^{\circ}$ for a chord-based Reynolds number $Re \equiv u_{\infty} c/{\nu}_{\infty}=400$ and Mach number $M_{\infty}\equiv u_{\infty}/a_{\infty}=0.1$. Here, $u_{\infty}$ is the free-stream velocity, $c$ is the chord length, ${\nu}_{\infty}$ is the kinematic viscosity, and $a_{\infty}$ is the freestream sonic speed.  
The DNS domain is shown in figure~\ref{para}(a), and the {\color{black}steady-state} vorticity field is presented in figure~\ref{para}(b).

\begin{figure}
  \centering
    \includegraphics[width=1\textwidth]{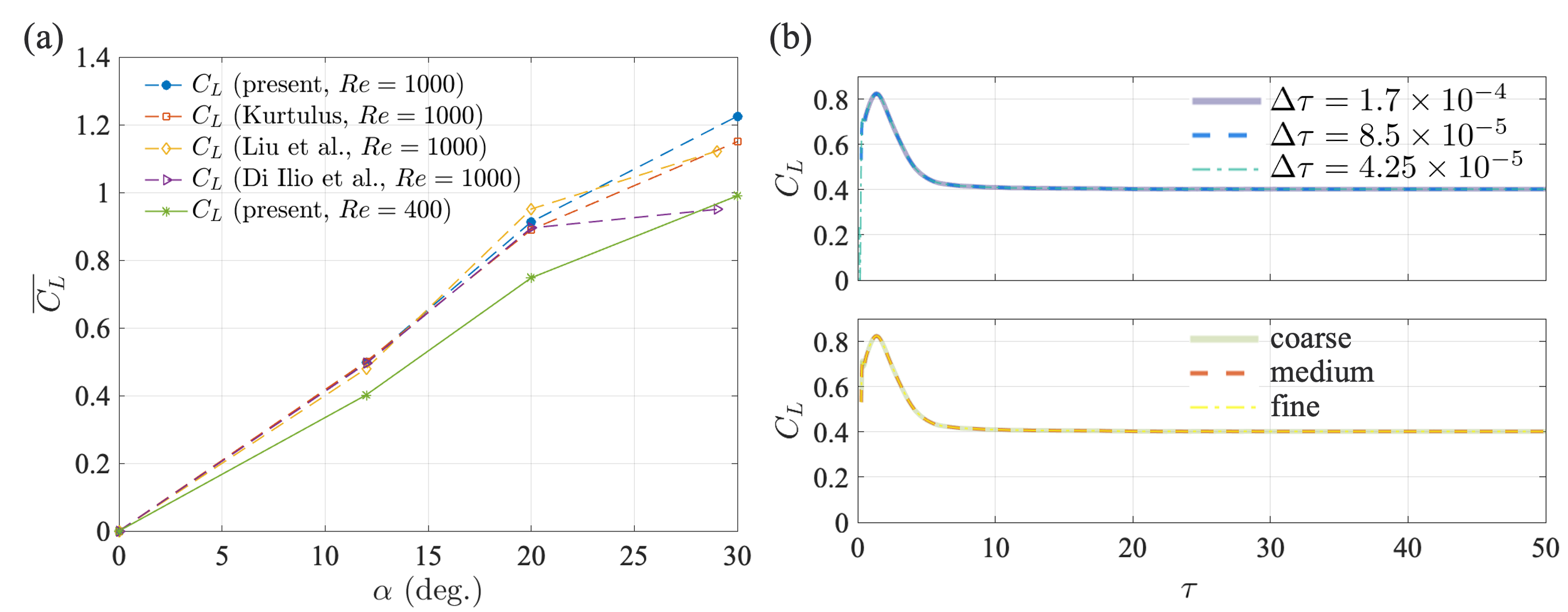}
    \caption{(a) Comparison of time-averaged lift coefficient between references and the present study over different angles of attack. 
    (b) Temporal and spatial convergence of lift history for a NACA0012 airfoil at the angle of attack $12^\circ$ and $Re=400$.}
    \label{validate}
\end{figure}

We validate the DNS model without a vortex gust by comparing the time-averaged lift $\overline{C_L}$ on a NACA0012 airfoil with previous studies~\citep{kurtulus2016wake,liu2012numerical,di2018fluid}, as shown in figure~\ref{validate}(a). At $Re=1000$, the $\overline{C_L}$ from our simulation agrees well with the references over various angles of attack. The $\overline{C_L}$ from $Re=400$ is generally lower than that from $Re=1000$.
In addition, time convergence and grid convergence are checked for our DNS results of the flow over a NACA0012 airfoil at $\alpha=12^\circ$ and $Re=400$. The coarse, medium, and fine meshes have grid points of 62566, 74046, and 194012, respectively. Figure~\ref{validate}(b) exhibits that both time and grid convergence are achieved for the lift coefficient $C_L$ over time. We chose the medium mesh and a CFL number of $a_{\infty}\Delta t/\Delta x<1$ in our DNS.
For the DNS of vortex-airfoil interactions, the results from the compressible solver CharLES has been compared to that from the incompressible solver Cliff. The lift coefficients match with each other with these two different solvers. In addition, time convergence and grid convergence are achieved for vortex-airfoil interactions.

%%%%%%%%%%%%%%%%%%%%%%%%%%%%%%%%%%%%
%%%%%%%%%%%%%%%%%%%%%%%%%%%%%%%%%%%%

%%%%%%%%%%%%%%%%%%%%%%%%%%%%%%%%%%%%
%%%%%%%%%%%%%%%%%%%%%%%%%%%%%%%%%%%%

A compressible Taylor vortex~\citep{taylor1918dissipation} is introduced upstream of the airfoil, whose angular velocity is given by
\begin{equation}
    u_{\theta}=u_{\theta \rm{max}}{\dfrac{r}{R}}{\rm exp}\biggl({\dfrac{1}{2}-\dfrac{r^2}{2R^2}}\biggr),
\end{equation}
where $r$ is the radial coordinate from the vortex center, $R$ is the radius of the vortex, $u_{\theta \rm{max}}$ is the maximum rotational velocity, and $(x_0,y_0)$ is the initial center position of the vortex, as shown in figure~\ref{para}(c)-(d). Here, the gust ratio is defined as
\begin{equation}
    G\equiv\frac{u_{\theta \rm{max}}}{u_{\infty}}.
\end{equation}
For the current study, we consider $G\in\{-1,-0.5,0.5,1\}$ and $R/c=0.5$. Disturbed flows with moderate ($G=\pm 0.5$) and strong ($G=\pm 1$) vortices are studied as representative examples of highly unsteady flow scenarios~\citep{zhong2023sparse, AnyaGust2022}. As the initial condition of the simulations, vortices are introduced with their centers at $(x_0,y_0)=(-3c,0)$. {\color{black}We define $\tau\equiv u_\infty t/c=0$ }as the time the center of the vortex arrives at the leading edge of the airfoil placed at $(x_{0},y_{0})=(0,0)$.

To start tracing the transient amplification of the fluid system with the optimal time-dependent modes, we need to choose an appropriate initial condition for the time-dependent modes and coefficient matrix. We take the initial guess of the perturbations $\textbf{\textsf{Q}}^{\prime}$ from the initial condition matrix 
\begin{equation}  \textbf{\textsf{Q}}^{\prime}_{0}\equiv[\textbf{\textsf q}^{\prime}_{\tau_a},\textbf{\textsf q}^{\prime}_{\tau_a+\Delta \tau},...,\textbf{\textsf q}^{\prime}_{\tau_b}]
        \in {\mathbb{R}}^{n\times {d}}, 
\end{equation}
where $\textbf{\textsf q}^{\prime}_{\tau}\equiv\textbf{\textsf q}_{\tau}-\textbf{\textsf q}_{\text{steady}}$, and $\textbf{\textsf q}_{\text{steady}}$ is the flow state matrix of steady state.
When the airfoil is at an angle of attack of $12^{\circ}$, a steady flow is achieved {\color{black}in the absence of} a vortical gust. For other flows without a steady state, $\textbf{\textsf q}_{\text{steady}}$ can be approximated by taking the time average of the flow.
The initial condition matrix is obtained by stacking $d$ snapshots within $\tau\in[\tau_a,\tau_b]$ with a constant time interval $\Delta\tau$. Then, the initial OTD modes and their coefficients are chosen from the singular value decomposition of the initial condition matrix $\textbf{\textsf{Q}}^{\prime}_{0}$ as
     \begin{equation}
        \textbf{\textsf{Q}}^{\prime}_{0}=\textbf{\textsf{U}}{\bm\Sigma} \textbf{\textsf{V}}^{\rm T}\approx\textbf{\textsf{U}}_r(\tau_0)\textbf{\textsf{Y}}_r^{\rm T}(\tau_0),
        \label{eq:perturb_decomp}
     \end{equation}
where $\textbf{\textsf{U}}\in {\mathbb{R}}^{n\times {d}}$ and $\textbf{\textsf{V}}\in {\mathbb{R}}^{{d}\times {d}}$ are the left and right singular vectors, respectively.
By choosing the leading $r$ vectors of $\textbf{\textsf{U}}$ as the initial time-dependent modes $\textbf{\textsf{U}}_r(\tau_0)$, the coefficient matrix $\textbf{\textsf{Y}}_r(\tau_0)$ can be initialized as the first $r$ vectors of $\textbf{\textsf{V}}\bm\Sigma$. Here, the initial evolution moment is indicated with $\tau_0$.

After obtaining the initial OTD modes and their coefficient{\color{black}s}, we use the fourth-order Runge-Kutta time-stepping method to solve the OTD evolution equations (\ref{eq:U_evolve})-(\ref{eq:Y_evolve}) with the time-dependent linear operator $\textbf{\textsf{L}}(t)$. To reduce the computational cost, the unsteady base state ${\bar{\textsf{\textbf q}}}(t)$ obtained from the DNS is interpolated into a smaller domain to extract the linear operator $\textbf{\textsf{L}}(t)$~\citep{sun2017biglobal}, as shown in figure~\ref{para}(a). The Dirichlet boundary condition is applied to the far-field boundary and the airfoil, and the Neumann boundary condition is set as the outlet boundary condition. With these boundary conditions, we extract the linear operator $\textbf{\textsf{L}}(t)\in {\mathbb{R}}^{n\times n}$ in a discrete form. Here, the degrees of the freedom of the discretized flow $n=428370$.
To further reduce the computational cost, we restrict the OTD domain as the linear operator domain overlapping with the rectangle {\color{black}$(x,y)/c\in [-4,8]\times [-2,2]$}. As all OTD modal structures appear around the airfoil, changing the OTD domain size does not affect the OTD modes and their coefficients.
 
In the current study, the initial time for OTD analysis is chosen as $\tau_0=-1$. The initial condition matrix is formulated by collecting {\color{black}a total of} $d=710$ flow state vectors from $[\tau_a,\tau_b]=[-1,-0.4]$, during which time the vortical gust advects from upstream of the airfoil toward the leading edge of the airfoil. 
The initial conditions are carefully selected before the vortex gust interacts with the airfoil. This choice is critical for analyzing the response to upstream disturbances and enabling timely interventions to suppress the growth of perturbations during the vortex-airfoil interaction.
 %{The initial condition is carefully chosen to consider the perturbation response when the airfoil encounters the upstream disturbance.}
We consider the overall OTD analysis over {\color{black}$\tau\in [-1,1]$}. Additional details on the initial modes are discussed in Appendix A, and discussions {\color{black}on} the optimal initial perturbation are presented in Appendix B.

%%%%%%%%%%%%%%%%%%%%%%%%%%%%%%%%%%%%
\subsection{Flow Physics}
\label{sec:physics}

%%%%%%%%%%%%%%%%%%%%%%%%%%%%%%%%%%%%
\begin{figure}
  \centering
    \includegraphics[width=1\textwidth]{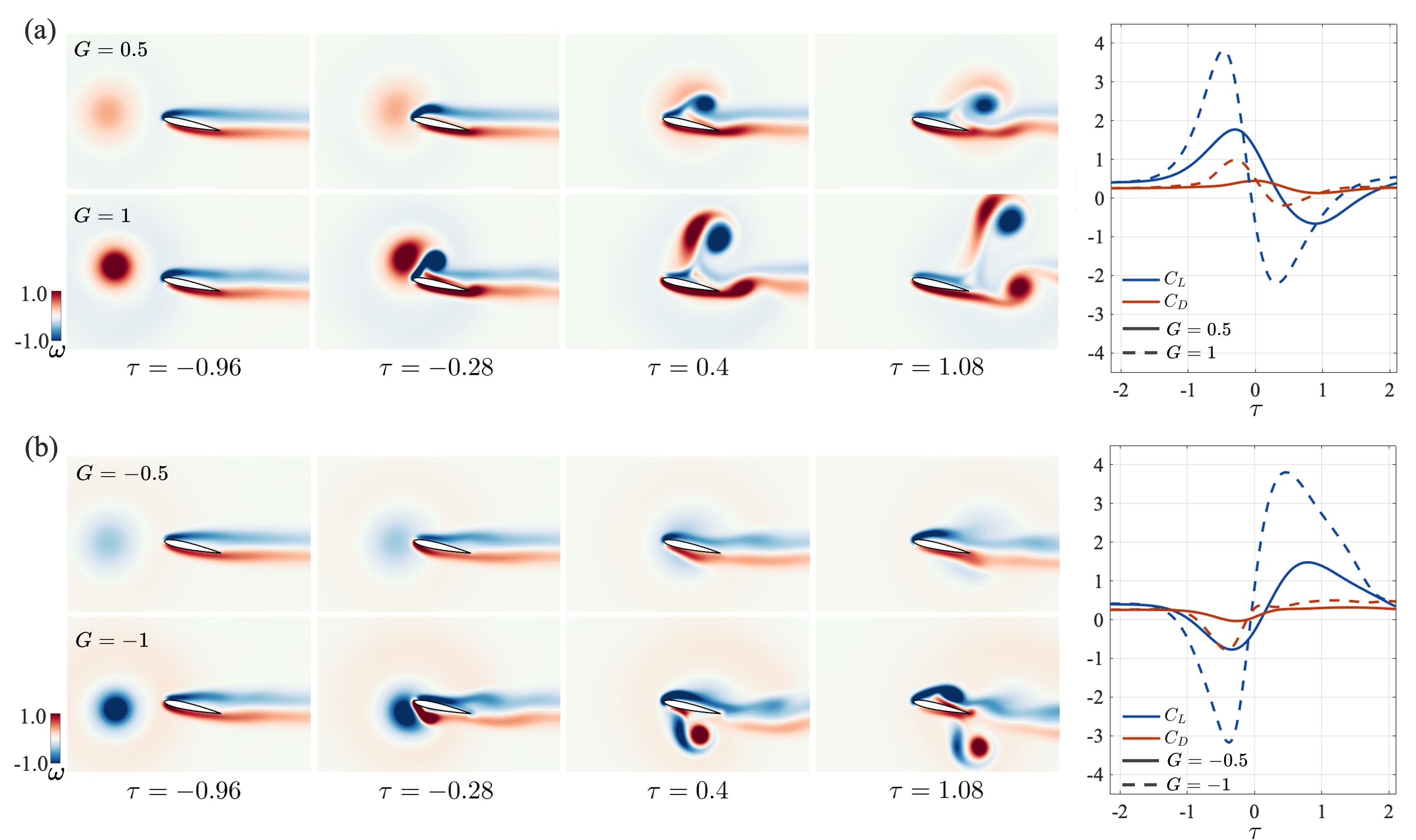}
    \caption{Vorticity fields and aerodynamic forces disturbed by a $(a)$~positive and $(b)$~negative vortex gust.}
    \label{physics}
\end{figure}
%%%%%%%%%%%%%%%%%%%%%%%%%%%%%%%%%%%%

Vortex-airfoil interactions exhibit strong transient characteristics within a short time. 
Previous studies have shown that vortex-body interaction may involve both rapid distortion of the incident vorticity field and injection of vorticity from the surface of the body~\citep{rockwell1998vortex, eldredge2019leading, martinez2020analysis, qian2022interaction, fukami2023grasping}.
To inform our interpretation of the dominant OTD modes, we analyze the perturbation evolution during the vortex-airfoil interaction process.
Here, we first analyze the dynamics of the unsteady base flows of the aforementioned four vortex-airfoil interaction cases. 
The time evolution of the vorticity fields and the aerodynamic forces of the vortex-airfoil interaction cases are presented in figure~\ref{physics}. 
The lift and drag coefficients are defined as $C_L= F_L /(0.5\rho u_{\infty}^2c)$ and $C_D=F_D /(0.5\rho u_{\infty}^2c)$, respectively, where $F_L$ is {\color{black}the lift force} and $F_D$ is {\color{black}the drag force} on the airfoil. Initially, the vortical disturbance is introduced upstream of the wing. This vortex approaches the airfoil and produces large {\color{black}transient effects} around the airfoil.

{\color{black}The} dynamics of a positive (counterclockwise) vortex interacting with a NACA0012 airfoil {\color{black}is presented in figure~\ref{physics}$(a)$}. A positive vortex gust induces lift and drag force increase as it impinges on the leading edge of the airfoil. Shortly thereafter, the aerodynamic forces decrease once the center of the vortex convects past the center of the airfoil. When a moderate positive vortex ($G=0.5$) first impinges on the airfoil, negative vorticity generated at the leading edge quickly rolls up into a {\color{black}leading-edge vortex (LEV)} above the airfoil surface, as seen at {\color{black}$\tau=-0.28$}. The increase of the negative circulation due to the growth of the LEV results in a lift augmentation on the airfoil~\citep{dickinson1993unsteady, eldredge2019leading}. After the LEV detaches from the leading-edge vortex sheet around ${\color{black}\tau}=1$, the vortex advects into the wake.
% The fast variations of the vorticity field indicate that the airfoil wake can undergo rapid distortions from the impingement of a positive vortical gust.   
For a strong positive vortex ($G=1$) shown in figure~\ref{physics}$(a)$, the interactions between the vortex and the airfoil generate a large LEV forming a vortex pair with the gust vortex, which moves far away from the airfoil body. Such a strong interaction produces four times larger lift and drag fluctuations compared to a moderate vortex-airfoil interaction with $G=0.5$.
% when the leading-edge vortex has a much larger vorticity magnitude and occupies a larger recirculation region near airfoil LE~\citep{golubev2011high}, as demonstrated in figure~\ref{physics}$(a)$ at ${\color{black}\tau}=-0.28$. 
In addition, a trailing-edge vortex is produced as the center of the strong vortex passes over the wing around ${\color{black}\tau}=1$.
The shedding of the trailing-edge vortex is not seen in the moderate positive vortex-airfoil interaction of $G=0.5$, indicating that the $G=1$ case has more drastic transient effects on wakes around the airfoil.

A negative (clockwise) vortex gust induces different effects on wake dynamics and aerodynamic performance compared to a positive vortex gust.
For a moderate negative vortex case ($G=-0.5$) shown in the upper row of figure~\ref{physics}$(b)$, the gust vortex does not cause large-scale flow separations as it convects around the airfoil{\color{black}, but imposes} lift variation of about 5 times the {\color{black}steady-state} lift. For a strong negative vortex gust of $G=-1$, such a disturbance induces massive flow separation on both sides of the wing, as shown in the bottom row of figure~\ref{physics}$(b)$. A large positive vortex is formed from the pressure-side roll-up around ${\color{black}\tau}=0.4$. When the disturbance moves away from the airfoil around ${\color{black}\tau}=1$, the remaining effect of the gust vortex is still sufficiently strong to disturb the flow on the suction side of the airfoil, producing a secondary leading-edge vortex.  
Moreover, the airfoil experiences lift reduction and augmentation that are about 10 times larger than the steady-state lift when interacting with a negative vortex gust.     
    
From these cases, we {\color{black}identify} different transient features of the disturbed airfoil wakes impinged by a gust vortex.  We consider these vortex-airfoil interaction cases as the unsteady base flows and examine the perturbation dynamics of each of these base flows using the OTD mode analysis below.

%%%%%%%%%%%%%%%%%%%%%%%%%%%%%%%%%%%%   
\section{OTD mode analysis}
\label{sec:results}
%%%%%%%%%%%%%%%%%%%%%%%%%%%%%%%%%%%%
\begin{figure}
  \centering
    \includegraphics[width=1\textwidth]{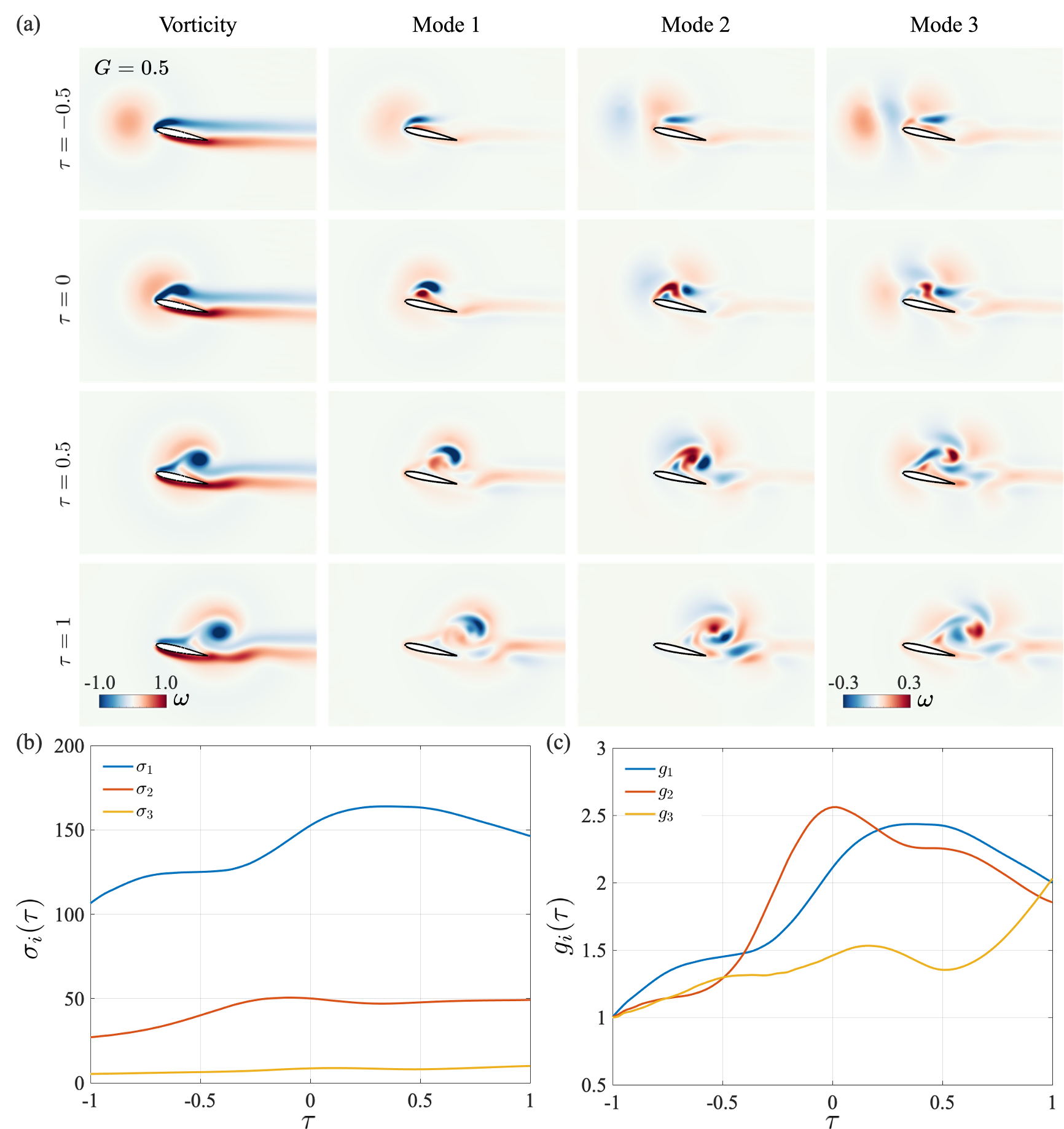}
    \caption{$(a)$~Vorticity fields of the time-varying base flow and the top three optimally time-dependent vorticity modes, $(b)$~the leading three singular values, and $(c)$~the leading three energy amplifications for $G=0.5$.
    }
    \label{Mpos_OTD}
\end{figure}
%%%%%%%%%%%%%%%%%%%%%%%%%%%%%%%%%%%%

Given the time-varying base flows studied in \S \ref{sec:physics}, the {\color{black}OTD} mode analysis is performed to reveal the transient flow structures that may be amplified throughout the vortex-airfoil interactions.  The OTD modes reveal regions where perturbations can undergo amplification with respect to the time-varying base flow.  Four interaction scenarios involving a moderate positive vortex ($G=0.5$), a strong positive vortex ($G=1$), a moderate negative vortex ($G=-0.5$), and a strong negative vortex ($G=-1$) are examined.  Verification of numerical calculations for the OTD {\color{black}modes} is provided in Appendix~\ref{Appen:verify}.

Let us first examine the {\color{black}the case of }moderate positive vortex ($G=0.5$), which advects over the airfoil and induces a new vortex roll-up above the airfoil. Given the transient variations of the unsteady base flow, we now investigate when and where perturbations can possibly be amplified through the lens of the OTD modes.
We present the leading three time-dependent modes in the order of {\color{black}their singular values $\sigma_i(\tau)$}. The vorticity fields and the top three time-dependent vorticity modes over time are shown in figure~\ref{Mpos_OTD}(a).  The temporal evolution of the leading three singular values is depicted in figure~\ref{Mpos_OTD}(b), and the leading three levels of energy amplification are shown in figure~\ref{Mpos_OTD}(c). 

Around ${\color{black}\tau}=-0.5$, the vortical gust advects toward the leading edge of the airfoil. The primary amplified region, highlighted by dominant spatial modes, appears around the top boundary layer near the leading edge of the airfoil. At the same time, the secondary amplified region emerges to coincide with the core of the vortical disturbance, shown upstream of the leading three OTD modes. Additionally, streamwise oscillations in the model structures appear for the higher-order modes 2 and 3. 
Therefore, during the period when the vortex approaches the airfoil, the primary amplified structure stems from the boundary layer near the leading edge of the airfoil and the subdominant sensitive regions correlate with the advection of the baseline vortical disturbance.

After the vortical gust impinges on the airfoil, the boundary layer separates and rolls up into an {\color{black}LEV}. Around ${\color{black}\tau}=0$, modes 1 and 2 identify the amplification of perturbations colocated and corotating with the LEV, as indicated above the suction side of the airfoil in figure~\ref{Mpos_OTD}(a). During the LEV formation, perturbations can experience maximum amplification along the core of the LEV. 
As the center of the vortical disturbance nears the half-chord position around the airfoil at ${\color{black}\tau}=0.5$, the LEV grows and moves toward the trailing edge of the airfoil. During this time, the most amplified region extends and shifts in the streamwise direction while rotating with the core of the LEV.
By ${\color{black}\tau}=1$, the LEV in the base flow has detached from the airfoil. During this time, the primary OTD mode still follows the advection and rotation of the LEV, while the wake region behind the trailing edge of the airfoil becomes increasingly important due to the trailing-edge vortex sheet roll-up. This shows the transition of the most amplified flow structure from the LEV to the wake behind the airfoil. 

By examining the singular values and energy amplifications, we identify the evolution of each OTD mode over time. Temporal changes of the corresponding singular values and energy amplifications $g_i(t)$ are presented in figures~\ref{Mpos_OTD}(b) and (c). The first OTD mode aligns with the most amplified direction during the evolution, associated with a large singular value of $\mathcal{O}(100)$. In figure~\ref{Mpos_OTD}(c), the maximum possible energy amplification $g_1(t)$ and the sub-optimal energy amplifications $g_2(t)$ and $g_3(t)$ have an increasing trend before {\color{black}$\tau=0$}. The increasing energy amplifications indicate that the perturbations can undergo a large transient growth as the LEV is forming. After the LEV detaches from the airfoil body around ${\color{black}\tau}=0.5$, the optimal and sub-optimal energy amplifications become twice their initial values, which uncovers that the perturbations can experience large amplification in the {\color{black}trailing-edge wakes}. 
Later, at ${\color{black}\tau}=1$, the third OTD mode shows growth in energy. This third OTD mode exhibits finer flow structures, indicative of its role in capturing more localized and higher-frequency dynamics within the flow field. As the interaction progresses, the observed increase in the energy growth of this mode suggests an amplification of these finer structures, likely driven by the development of small-scale instabilities and the transfer of energy from larger to smaller scales within the vortex-airfoil interaction.

% Compared with its initial value $\sigma_{1_0}$, the first singular value $\sigma_{1}(t)$ experiences a trend of first increase and then decrease, with a peak occurring around ${\color{black}\tau}=0.5$. After the peak, $\sigma_{1}/\sigma_{1_0}$ starts to decline, while $\sigma_{2}/\sigma_{2_0}$ and $\sigma_{3}/\sigma_{3_0}$ begins to grow, indicating that OTD modes 2 and 3 capturing the trailing-edge wake dynamics are increasingly important after ${\color{black}\tau}=0.5$.

%%%%%%%%%%%%%%%%%%%%%%%%%%%%%%%%%%%%
\begin{figure}
  \centering
    \includegraphics[width=1\textwidth]{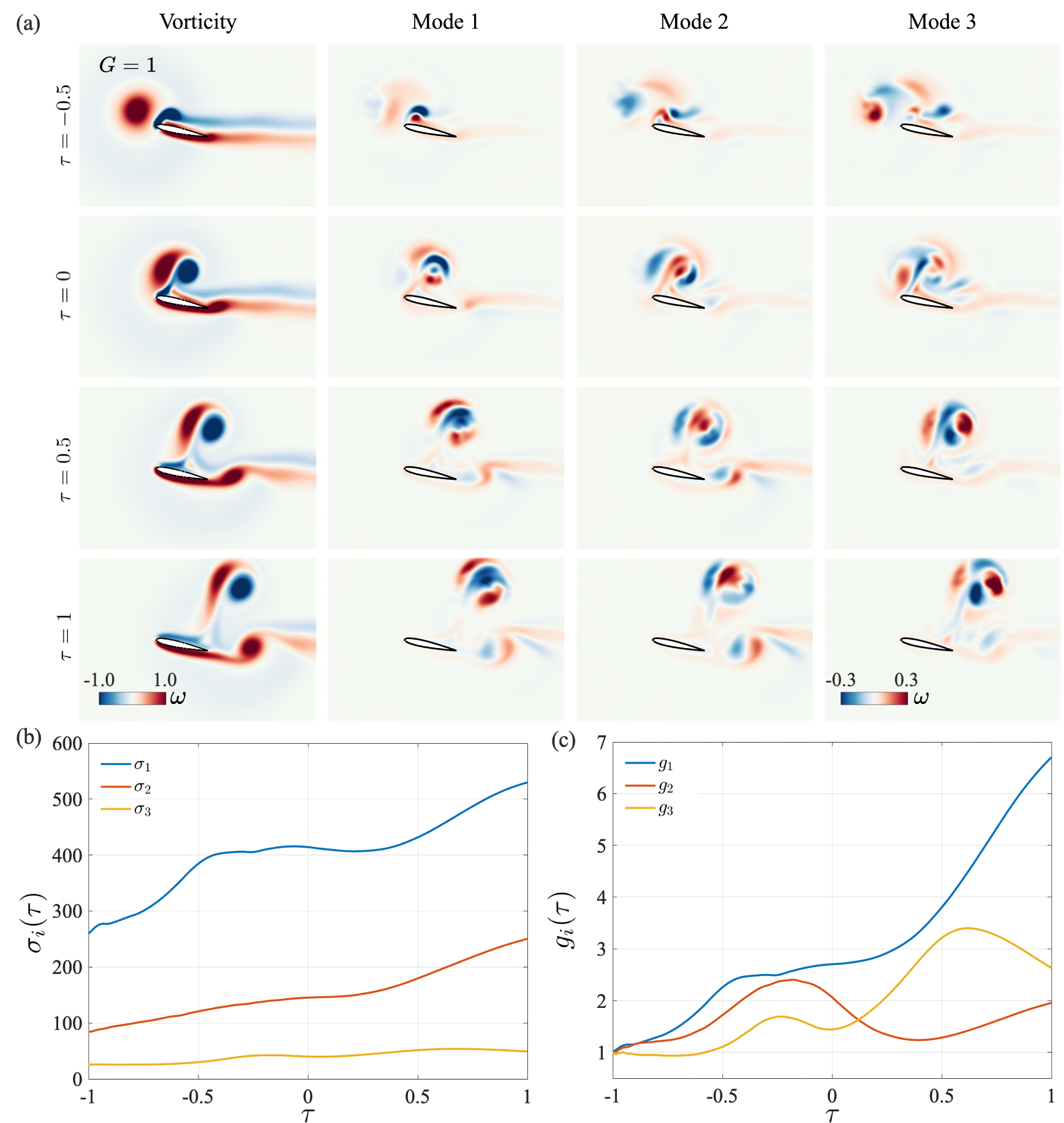}
    \caption{$(a)$~Vorticity fields of the time-varying base flow and the top three optimally time-dependent vorticity modes, $(b)$~the leading three singular values, and $(c)$~the leading three energy amplifications for $G=1$.}
    \label{Bpos_OTD}
\end{figure}
%%%%%%%%%%%%%%%%%%%%%%%%%%%%%%%%%%%%

%%%%%%%%%%%%%%%%%%%%%%%%%%%%%%%%%%%%
\begin{figure}
  \centering
    \includegraphics[width=1\textwidth]{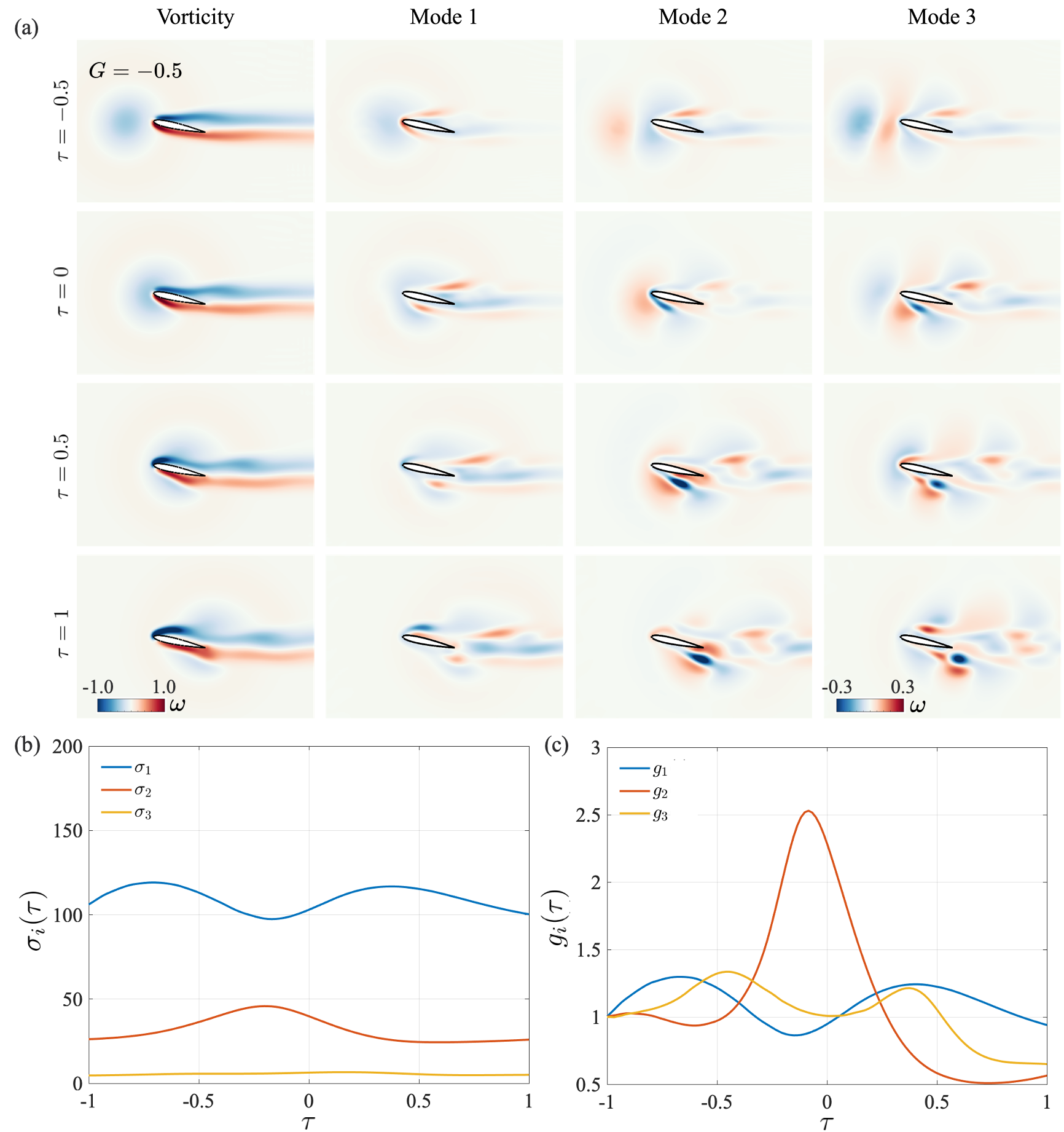}
    \caption{$(a)$~Vorticity fields of the time-varying base flow and the top three optimally time-dependent vorticity modes, $(b)$~the leading three singular values, and $(c)$~the leading three energy amplifications for $G=-0.5$.}
    \label{Mneg_OTD}
\end{figure}
%%%%%%%%%%%%%%%%%%%%%%%%%%%%%%%%%%%%

Now, let us consider the impact of vortex strength on the transient amplification of perturbations. In the case of a disturbed flow with a strong positive vortical gust of $G=1$, the vorticity field undergoes prominent transient fluctuations, as shown in the first column of figure~\ref{Bpos_OTD}(a).   
The dominant three spatial time-dependent modes are also presented. Around ${\color{black}\tau}=-0.5$, the strong vortex induces drastic deformation of the vortex sheet around the leading edge of the airfoil. During this time, the primary and secondary modes exhibit substantial amplification potential of perturbations in a localized region concentrated on the core of the vortex roll-up. Shortly after ${\color{black}\tau}=0$, the strong {\color{black}dipole (vortex pair)} forms over the airfoil surface. The core of the negative LEV emerges as the most amplified {\color{black}region} shown in modes 1 and 2. {\color{black}Mode 3 highlights a thin region between the positive vortex and the negative vortex in the LEV pair, suggesting that the stretching of vortex filaments within the LEV can lead to the growth of perturbations.}
     
{\color{black}Later,} when the vortex pair detaches from the airfoil after ${\color{black}\tau}=0.5$, all three dominant OTD modes identify the vortex pair as the most amplified flow structure. Concurrently, a small positive vortical structure in the base flow emerges at the trailing edge around ${\color{black}\tau}=0.5$. The amplified region behind the TE suggests that a secondary amplification of perturbations is associated with the formation of a TE vortex. 
From the analysis of the OTD modes over time, we find that the most amplified flow structures follow the vortices induced during the interaction between the strong vortical gust and the airfoil. {\color{black} The LEV pair, being a focal point of energy concentration, coincides with the maximum amplification regions of the OTD modes.}    
The dominant singular values and the energy amplifications over time are shown in figures~\ref{Bpos_OTD}(b)-(c). Across all three modes, singular values exhibit a steady increase from ${\color{black}\tau}=-1$ to $1$. The optimal energy amplification is approximately seven times its initial value, revealing that perturbations can be continuously amplified during the interaction between the vortex and the wing. {\color{black}The second and third energy growths are non-monotonic, suggesting that the higher-order modes have different temporal growth rates compared to the first mode.}

%%%%%%%%%%%%%%%%%%%%%%%%%%%%%%%%%%%%
\begin{figure}
  \centering
    \includegraphics[width=1\textwidth]{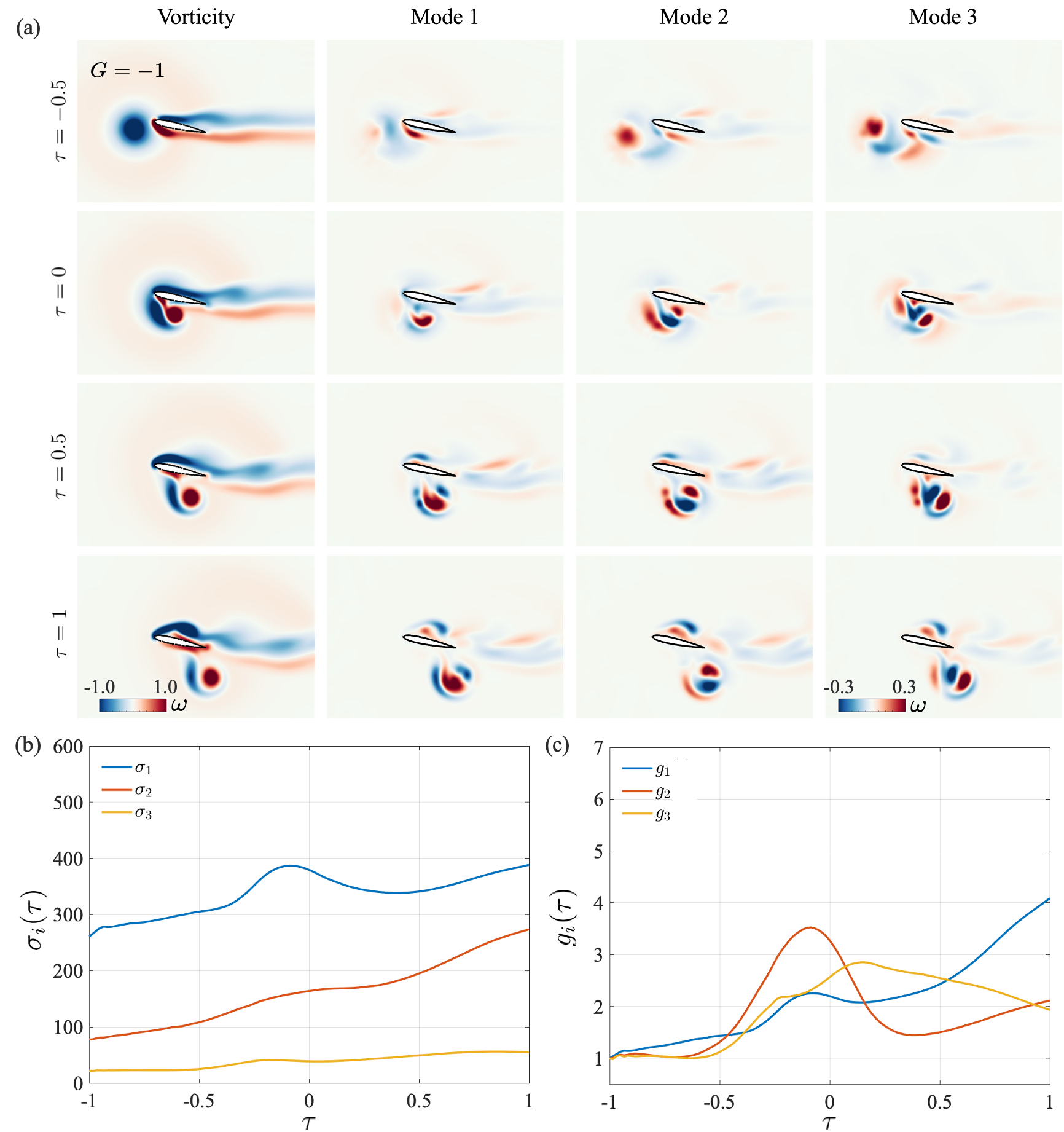}
    \caption{$(a)$~Vorticity fields of the time-varying base flow and the top three optimally time-dependent vorticity modes, $(b)$~the leading three singular values, and $(c)$~the leading three energy amplifications for $G=-1$.}
    \label{Bneg_OTD}
\end{figure}
%%%%%%%%%%%%%%%%%%%%%%%%%%%%%%%%%%%%

In contrast, when encountering a negative vortical gust, the dynamics of perturbations differ from those cases of positive vortex-airfoil interactions. We employ OTD mode analysis for the unsteady base flow concerning interactions between an airfoil and a moderate vortical gust ($G=-0.5$). Despite inducing significant lift and drag fluctuations on the airfoil, a moderate negative gust does not lead to large vortex shedding upon impingement, as visualized in figure~\ref{Mneg_OTD}(a).
Regarding the perturbation amplification, the primary OTD mode only exhibits a gradual change in the most amplified flow structures. The amplified regions identified by mode 1 are associated with the deformation of the wakes from ${\color{black}\tau}=-0.5$ to 1. 
 
{\color{black}The time evolution} of the amplified regions in modes 2 and 3 exhibits different perturbation dynamics from mode 1. Before ${\color{black}\tau}=-0.5$, modes 2 and 3 do not exhibit any prominent features in spatially sensitive areas. 
As the moderate negative vortex interacts with the airfoil around ${\color{black}\tau}=0$, the boundary layer separates from the pressure side of the airfoil and becomes highlighted by OTD modes 2 and 3. Subsequently, as the vortical gust advects downstream after ${\color{black}\tau}=0.5$, a compact amplified area from the bottom side vortex sheet roll-up is captured in the second and third modes. 
%This result suggests that there exists a small region near the bottom surface of the airfoil that is susceptible to perturbation for a moderate negative vortex-airfoil interaction. 
This result suggests that the second and third modes are more sensitive to localized disturbances and transient phenomena. This underscores the importance of higher-order modes in providing a comprehensive understanding of the vortex-airfoil interaction, particularly in capturing the nuanced flow features that govern the local perturbation behavior.

The corresponding singular values and perturbation amplifications are shown in figures~\ref{Mneg_OTD}(b)-(c). The leading singular value $\sigma_1=\mathcal{O}(100)$ remains fairly flat during its evolution, corresponding to the gradual changes of the leading OTD mode.
On the other hand, the secondary energy amplification $g_2(t)$ shows a spike around ${\color{black}\tau}=-0.1$, as depicted in figure~\ref{Mneg_OTD}(c). {\color{black}The large energy amplification of mode 2 indicates that the perturbations can grow large when the bottom side vortex sheet starts to roll up.} 
The ability of the second mode to detect and amplify during the onset of roll-up underscores its importance in tracking the timing of transient perturbation amplification.
{\color{black}Eventually all three energy amplifications become lower than 1 at $\tau=1$, suggesting a loss of coherent structures and a decrease in the energy of the OTD modes.}

{\color{black}Let us now consider the interactions between a strong negative vortical gust ($G=-1$) and an airfoil, which inherently include high levels of unsteadiness and nonlinearity.} As visualized in figure~\ref{Bneg_OTD}(a), a large vortex pair is generated from the pressure side of the airfoil around ${\color{black}\tau}=0$ during the violent interaction. After the vortex pair detaches from the airfoil body around ${\color{black}\tau}=1$, a smaller LEV is generated from the vortex sheet roll-up on the suction side of the airfoil. 
The three dominant modes in general highlight the most amplified region following the large vortex pair below the airfoil, shown in figure~\ref{Bneg_OTD}(a). Modes 1 to 3 have similar modal structures with notable features about the positive vortex. The most amplified structures coincide with the motion and rotation of the vortex pair that is generated around ${\color{black}\tau}=0$. In addition, a secondary sensitive region is found to be related to the subsequent vortex roll-up above the airfoil, as indicated by the emerging spatial structures about the LEV around ${\color{black}\tau}=1$.

In the figures~\ref{Bneg_OTD}(b)-(c), the variations of the three leading singular values and energy amplifications show that the amplification of perturbations generally becomes larger over $-1 \le {\color{black}\tau} \le 1$. In this case, the interaction between the strong negative vortex and the airfoil gives rise to a high level of {\color{black}perturbation amplifications}.
The sub-optimal energy amplification $g_2(t)$ exhibits one peak around ${\color{black}\tau}=-0.1$, as shown in the figure~\ref{Bneg_OTD}(c). During this time, both the positive and negative vorticity on the bottom side of the airfoil are highlighted by OTD mode 2. This result suggests that the perturbations can be amplified both in the LEV and the strong disturbance vortex. Later when the LEV convects into the wake near the airfoil trailing edge, the optimal energy amplification $g_1(t)$ becomes four times its initial value during evolution, as presented in figure~\ref{Bneg_OTD}(c). The increasing trend of energy growth indicates a substantial amplification of perturbations.

The uncovering of optimal time-dependent modes offers valuable insights for understanding transient amplification in response to perturbations about unsteady flows. By leveraging the most amplified structures discovered by OTD mode analysis, we can pinpoint the most amplified regions subject to time-varying perturbations. For a moderate positive vortex-airfoil interaction case ($G=0.5$), OTD modes unveil complex transitions of amplified flow structures from the leading-edge vortex sheet to the forming {\color{black}LEV}, with a secondary sensitive region as the wakes behind the airfoil. On the other hand, the leading amplified structures evolve slowly when the airfoil encounters a moderate negative vortical gust ($G=-0.5$). When the airflow is disturbed by a strong vortex ($G=\pm1$), OTD modes expose a direct correlation between shedding vortex trajectory and transient amplification of perturbations.

{\color{black}The energy amplification of OTD modes provides valuable insights into the underlying dynamics that govern the evolution of perturbations. A monotonic increase in energy amplification, as seen in $g_1(t)$ of $G=\pm 1$ cases, indicates a persistent instability mechanism. This mechanism for these two vortex-airfoil interactions is correlated with the high intensity of the vorticity of the primary shedding vortex. However, the region with the highest amplification is not always associated with the largest vorticity magnitude. For the example of $G=-0.5$, the highlighted modal structures of all three leading modes do not collapse on the leading edge of the airfoil where the highest vorticity magnitude is observed.
Diverse transient behaviors are uncovered by the variations of energy amplification. For example, $g_1$ and $g_2$ exhibit rapid increases followed by decreases, as shown in figure~\ref{Mpos_OTD}. A rapid increase in energy amplification indicates a highly unstable region of the flow. The roll-up of the LEV for $G=0.5$ corresponds to increasing energy amplifications. This region is prone to perturbations that can quickly amplify and lead to significant flow modifications. The subsequent decay suggests that the instability is transient, energy can be transferred downstream because of the wake formation.
}

{\color{black}OTD mode analysis can be used to identify perturbation dynamics for more complex flows, such as extreme aerodynamic flows and turbulent flows. Capturing the coherent structures that are receptive to perturbations in a time-evolving manner is critical in understanding the unsteady flow physics.
In addition, OTD mode analysis has the potential to guide flow control strategies.
OTD modes evolve dynamically, identifying when and where perturbations are most likely to grow. OTD modes inform when the flow is most receptive to perturbations, suggesting that actuators or energy inputs (e.g., actuation via jets or local forcing) can be applied at the most effective moments to amplify (or suppress) instabilities.
For example, for the $G=0.5$ case, when the vortex is approaching the airfoil, the most amplified region is the leading edge of the airfoil. The control effects at this moment should focus on the local region around the leading edge. On the other hand, when the vortex convects to the trailing edge of the airfoil, the trailing edge can be amplified more compared to the leading edge. A different local forcing near the trailing edge will be more efficient than the leading edge at this specific moment. }

% {Some implications proposed by Luke
% 1) It seems like OTD does pick out individual flow structures in a more intuitive way than POD, and it isn’t limited to single frequenct the way DMD is. Basically, it looks like OTD is a modal analysis technique where the modes reflect our intuition for the most important flow structures! That’s not an easy thing to find for a transient flow, so it may be worth stressing here that this identification of flow structures is not always straightforward with modal analysis, and OTD seems to align more with our intuition.  

% 2) OTD gives you a way of determining when optimal forcing/control transitions away from the shear layer to the wake. This is also usually a hard thing to determine, that OTD seems to do well. Note that Victoria is also evaluating this idea with resolvent analysis.

% 3) OTD basically gives you a criteria for which flow structures are “important” at each time in the flow. This could help with something like vortex panel methods, where only certain flow structures are modeled, and their diffusion is typically ad-hoc. I.e., OTD could help determine when to “remove” downstream vortices from a vortex simulation. Let me know if you’d like to see a reference or two on this idea. 

% 4) Are there any modeling implications here? That is, is anyone trying to formulate, say, a Galerkin model using OTD modes? If so, modeling could be worth mentioning here as a future effort/application.
% }

\section{Conclusions}
\label{sec:conclusion}

Vortex-airfoil interactions involve strong transient features, which are difficult to characterize with classical methods that are founded on time-invariant or periodic flow. In this study, optimally time-dependent (OTD) mode analysis was used to identify the most amplified flow structures of unsteady base flows of vortex-airfoil interactions. We considered four scenarios of vortex-airfoil interaction with a moderate positive vortex ($G=0.5$), a strong positive vortex ($G=1$), a moderate negative vortex ($G=-0.5$), and a strong negative vortex ($G=-1$). The optimally time-dependent modes capture the most amplified regions subject to perturbations, and the singular value variations offer insights into the significance of the OTD modes.
For the vortex-airfoil interaction of $G=0.5$, the amplified region transitions from the leading-edge vortex sheet to the evolving {\color{black}LEV}. In addition, a secondary amplified flow structure near the trailing-edge wake is revealed by subdominant OTD modes. When a strong positive vortical gust ($G=1$) interacts with an airfoil, perturbations are amplified following the advection and rotation of the {\color{black}LEV} pair generated from the impingement of the strong vortical gust. 
For a moderate negative vortex disturbance (G = -0.5), the flowfield exhibits few dominant vortical structures, and thus, the dominant OTD mode follows the advection of the wakes around the airfoil and undergoes gradual changes over time. For the strong negative vortex-airfoil interaction of $G=-1$, the shedding vortex pair convecting from the pressure side of the airfoil surface is identified as the most amplified flow structure subject to perturbations. 
The findings through the OTD mode analysis for the present vortex-gust airfoil interactions provide insights into the transient amplification of perturbations in unsteady flight conditions and hold promising implications for diverse unsteady flow problems.

% \newpage
\appendix
% \section{Appendix A: Verification of optimally time-dependent modes}
\section{Verification of optimally time-dependent modes}	
\label{Appen:verify}

\begin{figure}
      \centering
        \includegraphics[width=0.65\textwidth]{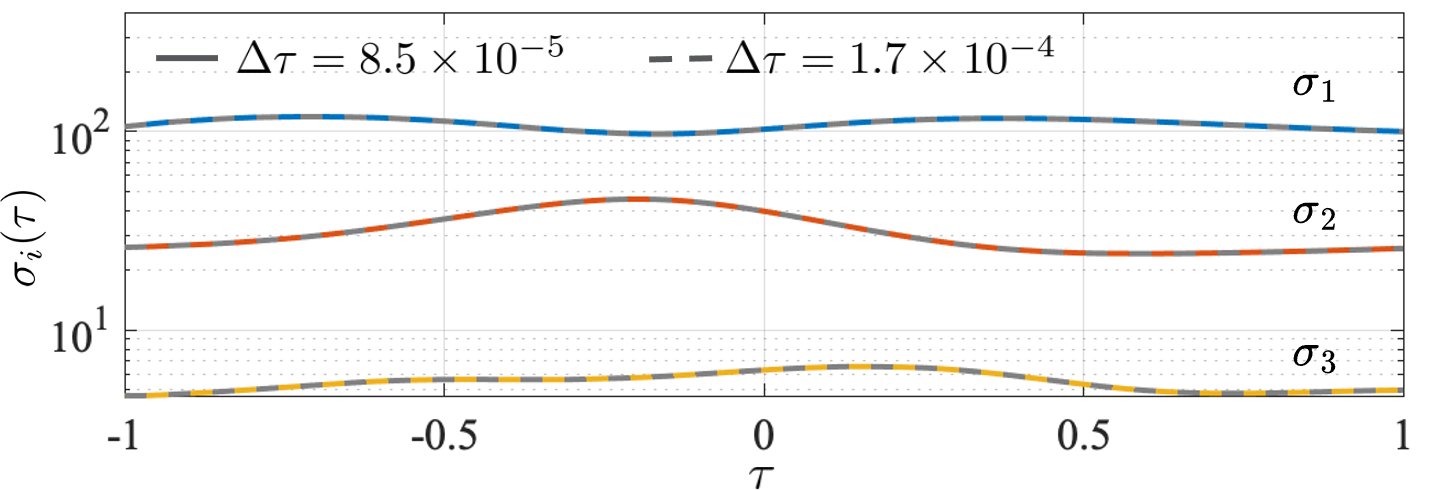}
        \caption{Time convergence on the top three singular values of moderate negative vortex-airfoil interaction. 
        }
        \label{time_converge}
    \end{figure}

The OTD modes are obtained by time integrating the evolution equation~\ref{eq:U_evolve}. In this appendix, we briefly outline the checks performed to ensure the accuracy of the OTD modes.  We first check the temporal convergence of the top three singular values, as shown in figure~\ref{time_converge} for the case of $G=-0.5$. Using different time step sizes of {\color{black}$\Delta \tau=8.5\times {10}^{-5}$} and $1.7\times {10}^{-4}$, the singular values from {\color{black}the} two cases match well with each other over time.  Therefore, we choose {\color{black}$\Delta \tau=1.7\times {10}^{-4}$} as the default time step for finding the optimally time-dependent modes.

    %%%%%%%%%%%%%%%%%%%%%%%%%%%%%%%%%%%%
    %%%%%%%%%%%%%%%%%%%%%%%%%%%%%%%%%%%%  

\begin{figure}
  \centering
    \includegraphics[width=1\textwidth]{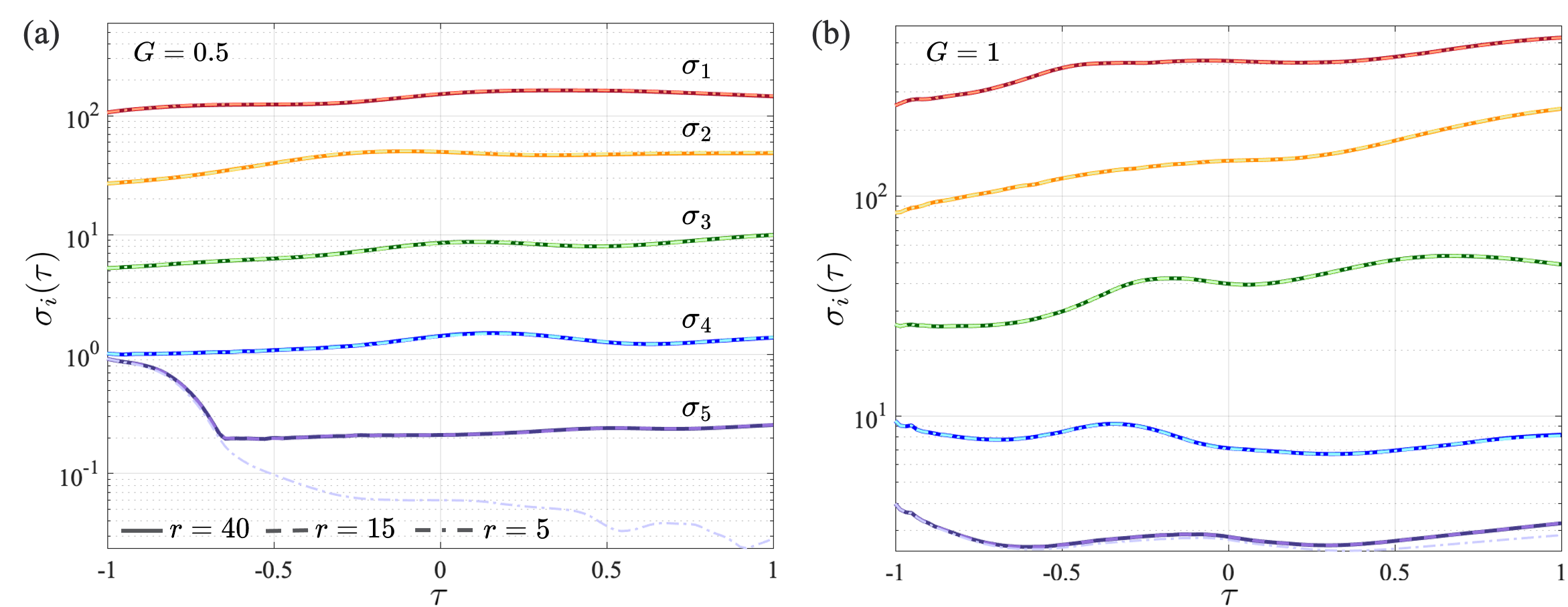}
\caption{Convergence on the number of OTD modes for the top five singular values of four disturbed flow cases:~$(a)$~$G=0.5$, $(b)$~$G=1$.}
    \label{sigCY_converge}
\end{figure}

Next, we examine the convergence on the number of OTD modes $r$. This is studied by comparing the singular values for different numbers of modes $r$.
For brevity, we present only positive vortex-airfoil interaction cases in the following analyses. The negative vortex-airfoil interaction cases exhibit analogous results.
As presented in figure~\ref{sigCY_converge} for positive gust cases, the top five singular values over time are shown for $r=5, 15$, and $40$. 
The solid lines represent the $r=40$ cases, the dashed lines denote the $r=15$ cases, and the {\color{black}dashed-dot} lines are the $r=5$ cases. 
In general, the leading three singular values of cases with different numbers of modes agree perfectly with each other. However, the lowest $r=5$ cases have discrepancies in the fourth and fifth singular values compared to cases with a larger number of modes.
Hence, we concluded that the singular values exhibit convergence with 15 modes for positive vortex-airfoil interaction cases.

%%%%%%%%%%%%%%%%%%%%%%%%%%%%%%%%%%%%
%%%%%%%%%%%%%%%%%%%%%%%%%%%%%%%%%%%% 

\begin{figure}
  \centering
    \includegraphics[width=0.6\textwidth]{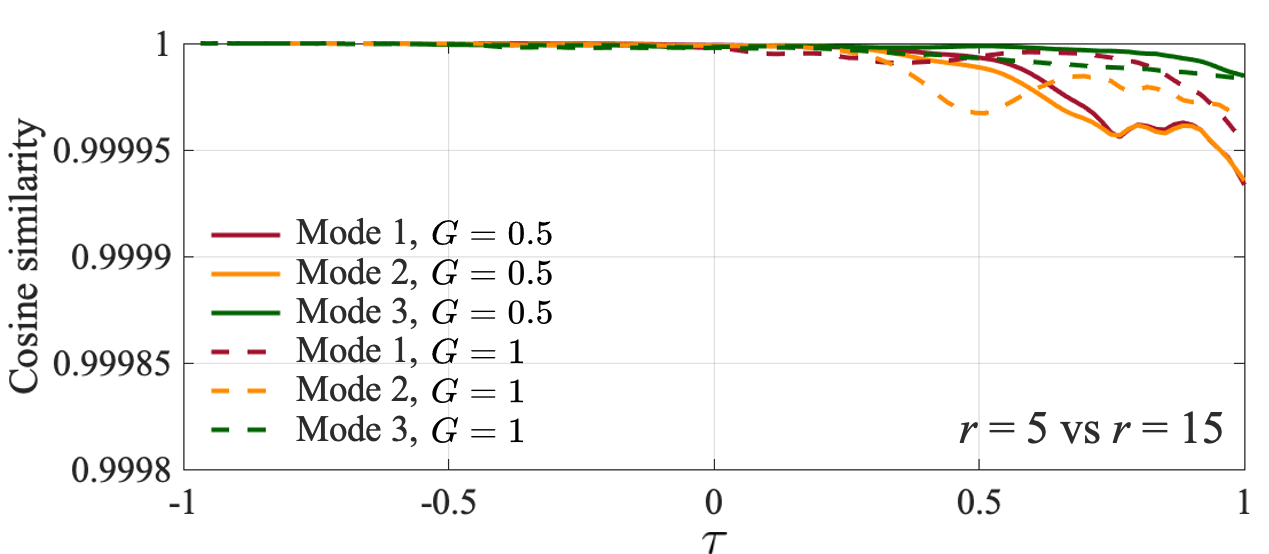}
\caption{Cosine similarity of each of the three dominant modes between $r=5$ and $r=15$ of positive vortex-airfoil interaction cases.}
    \label{simi_converge}
\end{figure}

The convergence check on the number of modes is also performed by comparing the cosine similarity between modes. 
Cosine similarity is defined as the inner product of two vectors, whose absolute value is in the range of $[0,1]$. When the absolute cosine similarity approaches 1, it indicates that the two vectors are similar to each other. For convenience, we present the absolute value of the cosine similarity between OTD modes below.
We examine the cosine similarity of the dominant three modes between $r=5$ and $15$ cases. Figures~\ref{simi_converge}(a) and (b) show the vortex-airfoil interaction cases for positive vortices, respectively. Among the comparisons of positive vortex cases of $G=0.5$ and $1$, all three modes from the $r=5$ cases achieve over 99.99$\%$ cosine similarity with the corresponding modes from $r=15$ cases. The high similarity levels demonstrate the high {\color{black}agreement} of the leading three modes extracted from disturbed flow scenarios. With only $r=5$ modes, the top three modes are as accurate as using 15 modes.

%%%%%%%%%%%%%%%%%%%%%%%%%%%%%%%%%%%%
%%%%%%%%%%%%%%%%%%%%%%%%%%%%%%%%%%%% 

%%%%%%%%%%%%%%%%%%%%%%%%%%%%%%%%%%%%
\begin{figure}
  \centering
    \includegraphics[width=1\textwidth]{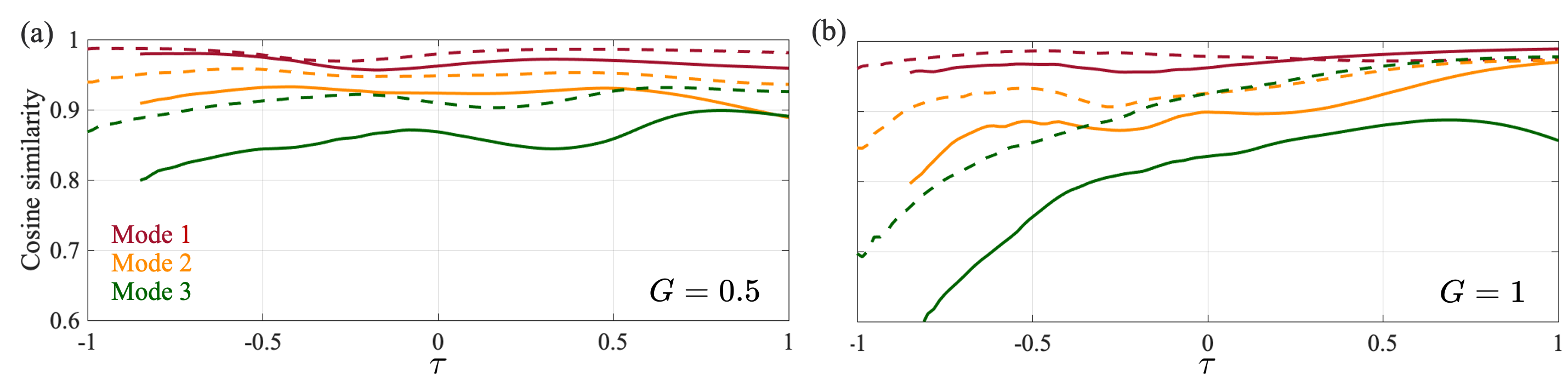}
    \caption{The influence of initial time for OTD evolution. For each of the leading three OTD modes, cosine similarity is checked between {\color{black}modes computed with initial time} $\tau_0=-0.85$ and $-1$ (solid lines), and between $\tau_0=-1.12$ and $-1$ (dashed lines). The initial condition matrix is extracted from the flow state snapshots over $[\tau_a,\tau_b]\in[-1,-0.4]$.}
    \label{start}
\end{figure}
%%%%%%%%%%%%%%%%%%%%%%%%%%%%%%%%%%%%   
    
{\color{black}Now we investigate the influence of initial evolution time $\tau_0$ on the OTD modes, the leading singular values converge while being insensitive to the choice of the OTD initial condition time $\tau_0$.
Using the same initial condition matrix considering $[\tau_a,\tau_b]=[-1,-0.4]$, we integrate the initial OTD modes from different initial times of $\tau_0=-1.12$, $-1$, and $-0.85$.}
The influence of different initial times on the evolution of the leading three modes is presented in figure~\ref{start}. 
The solid lines indicate the comparison of each of the leading three modes {\color{black}computed with} $\tau_0=-0.85$ and $-1$, and the dashed lines are {\color{black}computed with} $\tau_0=-1.12$ and $-1$. 
{\color{black}In the case of $G=0.5$, each of the three dominant modes with different initial evolution moments maintains over 80$\%$ similarity throughout the entire time span.}
For the vortex-airfoil interaction cases of $G=1$, however, the similarities start from relatively low values and increase afterward. 
{\color{black}This indicates that the observed dominant features converge to the same subset of OTD modes regardless of the initial evolution moments.}
% Note that in figure~\ref{start}(d), a sharp transition of cosine similarity is observed for modes 2 and 3, as a result of mode-switching.   
Based on these results, we perform the OTD mode evolution from $\tau_0=-1$, and the initial condition is given by $[\tau_a,\tau_b]=[-1,-0.4]$.

\begin{figure}
  \centering
    \includegraphics[width=1\textwidth]{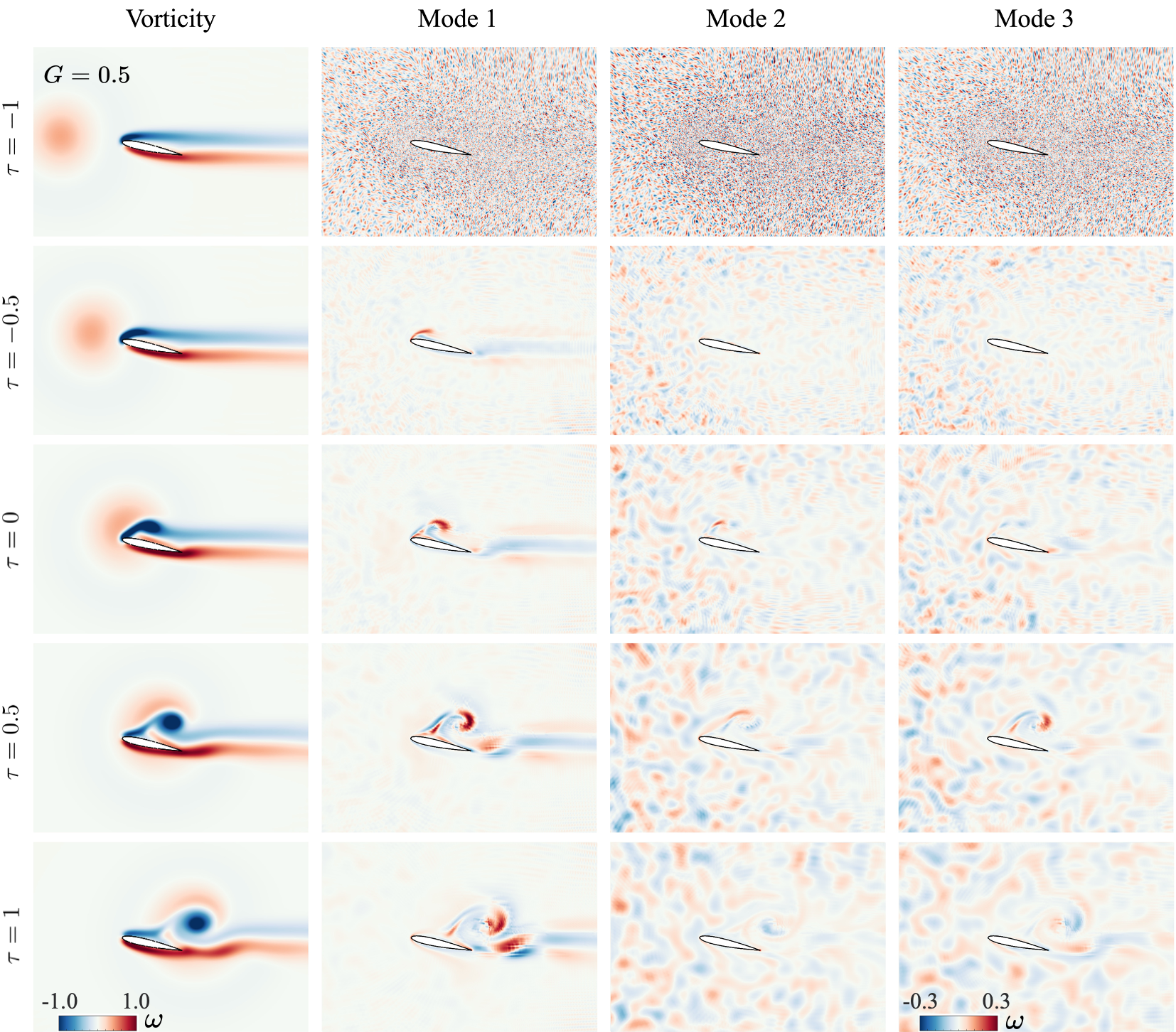}
    \caption{\color{black}The evolution of OTD modes with random noise as the initial condition, $G=0.5$.}
    \label{noise}
\end{figure}
%%%%%%%%%%%%%%%%%%%%%%%%%%%%%%%%%%%%   

% %%%%%%%%%%%%%%%%%%%%%%%%%%%%%%%%%%%%
% \begin{figure}
%   \centering
%     \includegraphics[width=0.7\textwidth]{Figs/sigma_noise.png}
%     \caption{\color{black}The leading three singular values of OTD modes with random noise as the initial condition, $G=0.5$.}
%     \label{sig_noise}
% \end{figure}
% %%%%%%%%%%%%%%%%%%%%%%%%%%%%%%%%%%%%  

{\color{black}To show that the primary OTD mode shown earlier in figure~\ref{Mpos_OTD} is not dependent on the initial condition, let us also consider random noise as the initial condition and evolve the OTD modes in time for the $G=0.5$ case. We use 15 orthonormal random noise vectors as the initial OTD modes at ${\color{black}\tau}=-1$. 

As shown in the figure~\ref{noise}, the OTD mode 1 captures very similar structures to the case shown in the figure~\ref{Mpos_OTD} when ${\color{black}\tau}>0$. The amplified region starts from the leading edge vortex sheet around ${\color{black}\tau}=-0.5$, then evolves with the {\color{black}shedding of the LEV}. Later around ${\color{black}\tau}=1$, the region after the trailing edge of the airfoil is highlighted as the most amplified region. On the other hand, OTD modes 2 and 3 are filled with random noises evolved from the initial condition. The noise structures get smoothed out as time increases, and vortical structures similar to OTD mode 1 are observed near the airfoil. 
% The leading three singular values of OTD modes are also presented in figure~\ref{sig_noise}. As $\sigma_1$ has the order of 1, $\sigma_2$ and $\sigma_3$ only have the order of $10^{-8}$, indicating that the dominance of first OTD mode is much larger than the higher-order modes.

Compared to the initial modes, which are the singular vectors from the initial perturbation matrix, random noise as the initial condition yields a worse result in capturing the perturbation amplification. Although the OTD modes converge to the most amplified structures subject to perturbations, the convergence requires a large number of initial condition vectors.
Therefore, we infer that a proper set of initial OTD modes can be selected from the SVD of the initial condition matrix collected around the OTD evolution starting time.}

% \section*{Appendix B: Most amplified initial perturbation}
\section{Most amplified initial perturbation}
\label{Appen:opt_pert}

After acquiring knowledge about transient amplifications from spatial modes, we can identify which perturbation has the maximum impact on the disturbed flows.
{\color{black}This information is crucial for understanding the underlying mechanisms driving flow instabilities and transient growth phenomena.}
The optimal initial perturbation that leads to the maximum amplification at time {\color{black}$\tau^*$} is obtained via Eq.~\ref{eq:inti_opt}.
Starting from $\tau_0$ to $\tau^*$, the most amplified initial perturbation can be written as 
\begin{equation}\label{eq:opt_ic}
    {\textbf{\textsf{q}}}_0^{\prime *}=\hat{\textbf{\textsf{U}}}_r(\tau_0){\rm\bm \Sigma}_r(\tau_0)\hat{\textbf{\textsf{Y}}}_r^{\rm T}(\tau_0)\hat{\textbf{\textsf{y}}}_1(\tau^{*})\in  {\mathbb{R}}^{n},
\end{equation}
which leads to a largest singular value at $\sigma_1(\tau^*)$ by evolving ${\textbf{\textsf{q}}}_0^{\prime *}$ in OTD subspace from $\tau_0$ to $\tau^*$. {\color{black} It is important to note that the optimal initial condition determined via equation \ref{eq:opt_ic} is confined to the space spanned by the columns of $\textbf{\textsf{Q}}'_0 \approx \hat{\textbf{\textsf{U}}}_r(\tau_0){\rm\bm \Sigma}_r(\tau_0)\hat{\textbf{\textsf{Y}}}_r^{\rm T}(\tau_0)$. }
\begin{figure}
  \centering
    \includegraphics[width=1\textwidth]{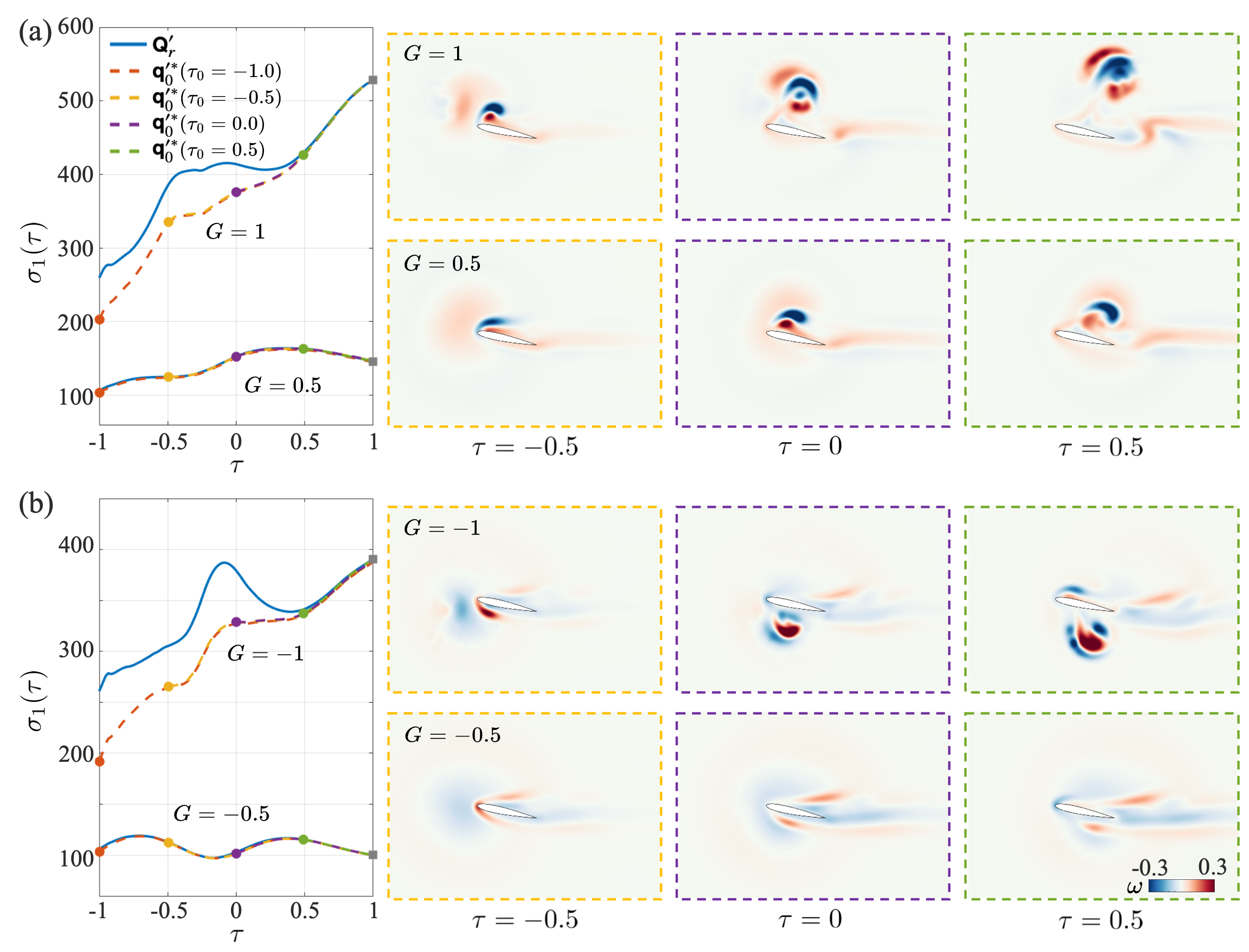}
    \caption{Evolution of the leading singular value subject to different ${\textbf{\textsf{q}}}_0^{\prime *}$ at $\tau_0 = -1, -0.5, 0,$ and $0.5$ (\text{shown with circles}), $\tau^*=1$ (\text{square}). The vorticity fields surrounded by yellow, purple, and green dashed boxes indicate the most amplified initial perturbations at $\tau_0= -0.5, 0,$ and 0.5, respectively. {\color{black}Each perturbation is normalized for visualization.}}
    \label{optP_start}
\end{figure}
%%%%%%%%%%%%%%%%%%%%%%%%%%%%%%%%%%%% 

To show the most amplified perturbation at different $\tau_0$ that arrives at the largest singular value at $\tau^*=1$, we compare its singular value with the largest singular value from the evolution of the set of perturbations ${\textbf{\textsf{Q}}}_r^{\prime}\equiv\hat{\textbf{\textsf{U}}}_r(\tau_0=-1){\rm\bm \Sigma}_r(\tau_0=-1)\hat{\textbf{\textsf{Y}}}_r^{\rm T}(\tau_0=-1)$ in the OTD subspace. The product of the OTD modes and the associated coefficients approximates the evolution of the set of perturbations ${\textbf{\textsf{Q}}}_r^{\prime}$, which is referred to as the baseline case in this section. We present the leading singular value of evolving ${\textbf{\textsf{Q}}}_r^{\prime}$ over time with blue solid lines in figure~\ref{optP_start}. 
Starting from four initial conditions at $\tau_0= -1, -0.5, 0,$ and $0.5$, the most amplified initial perturbations are identified from ${\textbf{\textsf{q}}}_0^{\prime *}=\hat{\textbf{\textsf{U}}}_r(\tau_0){\rm\bm \Sigma}_r(\tau_0)\hat{\textbf{\textsf{Y}}}_r^{\rm T}(\tau_0)\hat{\textbf{\textsf{y}}}_1(\tau^{*}=1)$. The singular value variations are shown on the left-hand side of figure~\ref{optP_start}. For the vortex-airfoil interaction cases of $G=\pm0.5$, the singular values from ${\textbf{\textsf{q}}}_0^{\prime *}(\tau_0= -1, -0.5, 0,$ and $0.5)$ almost collapse with the leading singular value of evolving the full set of perturbations ${\textbf{\textsf{Q}}}_r^{\prime}$. This result reveals that there exists one single perturbation that can be amplified the most over time. On the other hand, it is observed that the most amplified initial perturbations from cases of $G=\pm1$ have lower singular values than the largest singular value of the baseline case. {\color{black} This suggests that a strong vortex-airfoil interaction }
The normalized ${\textbf{\textsf{q}}}_0^{\prime *}$ subject to different $\tau_0$ are shown on the right-hand side of figure~\ref{optP_start}. For all interaction cases, they follow a similar shape of the leading mode $\hat{\textbf{\textsf{u}}}_1(t)$, indicating that the regions in which perturbation is amplified the most can be captured by the leading OTD mode.
      
%%%%%%%%%%%%%%%%%%%%%%%%%%%%%%%%%%%%
\begin{figure}
  \centering
    \includegraphics[width=1\textwidth]{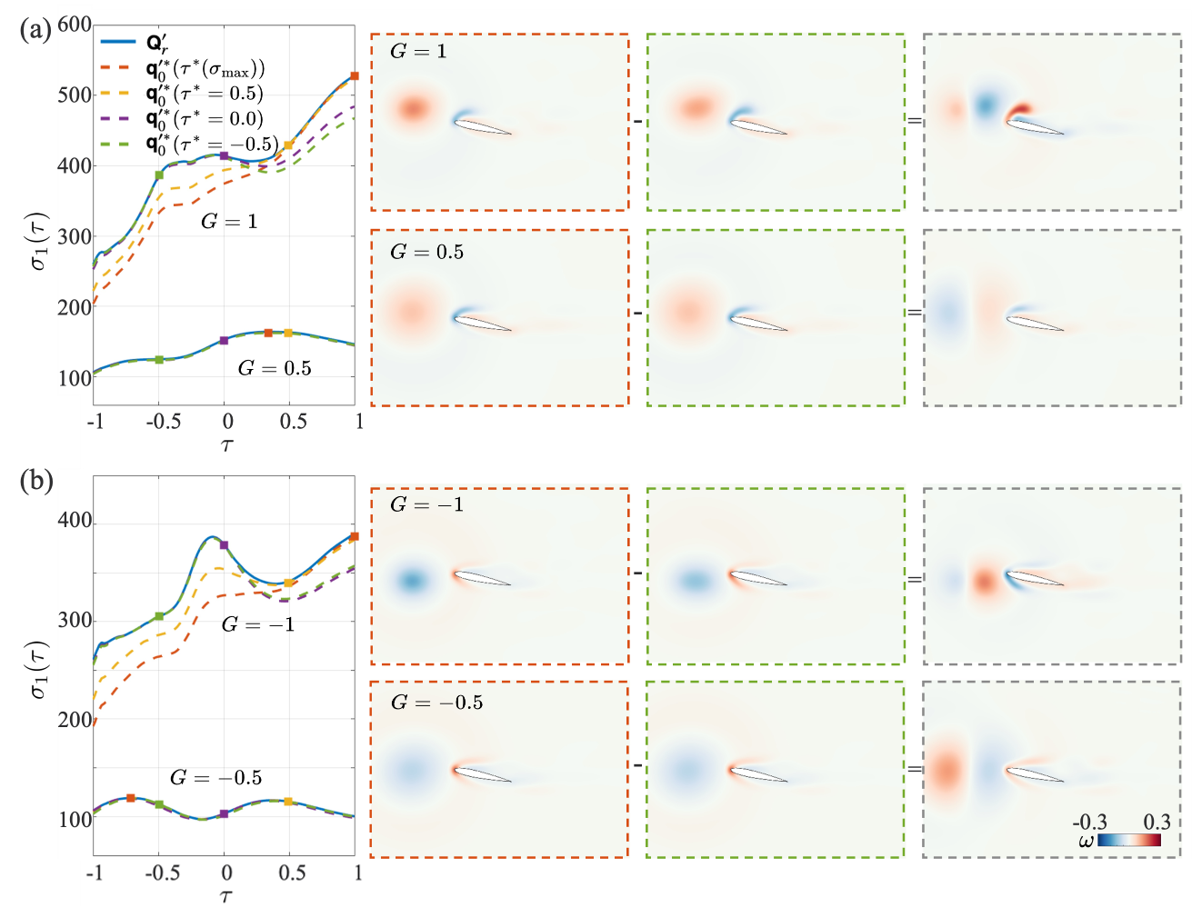}
    \caption{Evolution of the leading singular value subject to different ${\textbf{\textsf{q}}}_0^{\prime *}$ at $\tau^* = \tau^{*}(\sigma_{\rm{max}}), -0.5, 0,$ and $0.5$ (\text{denoted as squares}) with $\tau_0=-1$. The vorticity fields surrounded by orange, green, and gray dashed boxes indicate the most amplified perturbations at $\tau^*= \tau^*(\sigma_{{\rm{max}}}), -0.5,$ and the difference between them, respectively. {\color{black}Each perturbation is normalized for visualization.}}
    \label{optP_end}
\end{figure}
%%%%%%%%%%%%%%%%%%%%%%%%%%%%%%%%%%%%  
    
{\color{black}Moreover, figure~\ref{optP_end} presents the most amplified perturbations for a fixed $\tau_0$ and different $\tau^*$.} The initial evolution time are the same at $\tau_0=-1$, and the target singular values are at $\tau^*=\tau^{*}(\sigma_{\rm{max}}),-0.5,0,$ and $0.5$. By comparing the difference between initial perturbations at the same $\tau_0$, we are able to identify the key features of the most dangerous initial perturbation that can be amplified the most over the whole interaction period.
For moderate vortex-airfoil interaction cases as presented in figure~\ref{optP_end}, the singular value variations generally have the same trend as the baseline case. This indicates that the identified ${\textbf{\textsf{q}}}_0^{\prime *}(\tau_0=-1)$ subject to different $\tau^*$ are similar to each other, which can also be visualized on the right-hand side of figure~\ref{optP_end}. 
For strong vortex-airfoil interactions, however, the identified ${\textbf{\textsf{q}}}_0^{\prime *}(\tau_0=-1)$ that leads to a largest singular value of $\tau^*\geq0.5$ possesses a lower singular value than the baseline case. 
In figure~\ref{optP_end}, we compare normalized ${\textbf{\textsf{q}}}_0^{\prime *}(\tau^*(\sigma_{\rm{max}}))$ with normalized ${\textbf{\textsf{q}}}_0^{\prime *}(\tau^*=-0.5)$. For strong vortex-airfoil interactions ($G=\pm1$), ${\textbf{\textsf{q}}}_0^{\prime *}(\tau^*(\sigma_{\rm{max}}))$ has a higher intensity of vorticity at vortex center, and ${\textbf{\textsf{q}}}_0^{\prime *}(\tau^*=-0.5)$ has a sparsely distributed vorticity region in the upstream.    
This result indicates that a concentrated perturbation on the vortex core upstream is likely to be amplified more than a sparsely distributed perturbation.

{\color{black}In summary, for a time-varying fluid system, the most amplified initial perturbation within the initial perturbation subspace   evolves dynamically with time. Identifying these time-dependent, most amplified perturbations through OTD mode analysis highlights the key structures that have the potential to drive significant deviations from the base flow. This understanding is invaluable for perturbation amplification analysis of unsteady fluid systems.}

\section*{Acknowledgments}
\label{sec:acknowledgments}

YZ and KT acknowledge the support from the US Air Force Office of Scientific Research (Grant Number: FA9550-21-1-0178) and the US Department of Defense Vannevar Bush Faculty Fellowship (Grant Number: N00014-22-1-2798).

\section*{Declaration of interest}
\label{sec:doi}
The authors report no conflict of interest.

% \bibliography{taira_refs,Hessam}

\bibliographystyle{jfm}

% \section*{Outline}
% \begin{itemize}
%     \item Abstract
%     \item Introduction
%     \item Methodology: OTD
%     \item Problem set-up: physics and POD modes of 2 cases
%     \item OTD verification and convergence: dt convergence; Convergence on the number of modes
%     \item Results
%     \begin{itemize}
%         \item Transient features:
%         \begin{itemize}
%             \item OTD modes capture the shear-layer instability as the primary instability of vortex-airfoil interactions. The Kelvin-Helmholtz instability downstream can be predicted as the secondary instability with a positive vortex gust while the emergence of a near-steady-state wake mode is predicted for a negative vortex-airfoil interaction.            
%         \end{itemize}
%     \end{itemize}
%     \item Conclusion
% \end{itemize}

\end{document}